\input{aipcheck}

\documentclass[
    ,final            
  ,numberedheadings 
  ]
  {aipproc}

\layoutstyle{6x9}

\begin{document}

\title{Strangeness observables and pentaquarks}

\author{Sonia Kabana}{
  address={Laboratory for High Energy Physics, University of Bern,
 Sidlerstrasse 5, 3012 Bern, Switzerland}
}

\begin{abstract}
We review the experimental evidence on firstly, strangeness 
production as a signature for the QCD phase transition
and secondly, pentaquarks, the latest and most exotic manifestations of strangeness
in hadrons.
\end{abstract}

\maketitle

\section{Introduction}

\begin{figure}
  \includegraphics[height=.32\textheight]{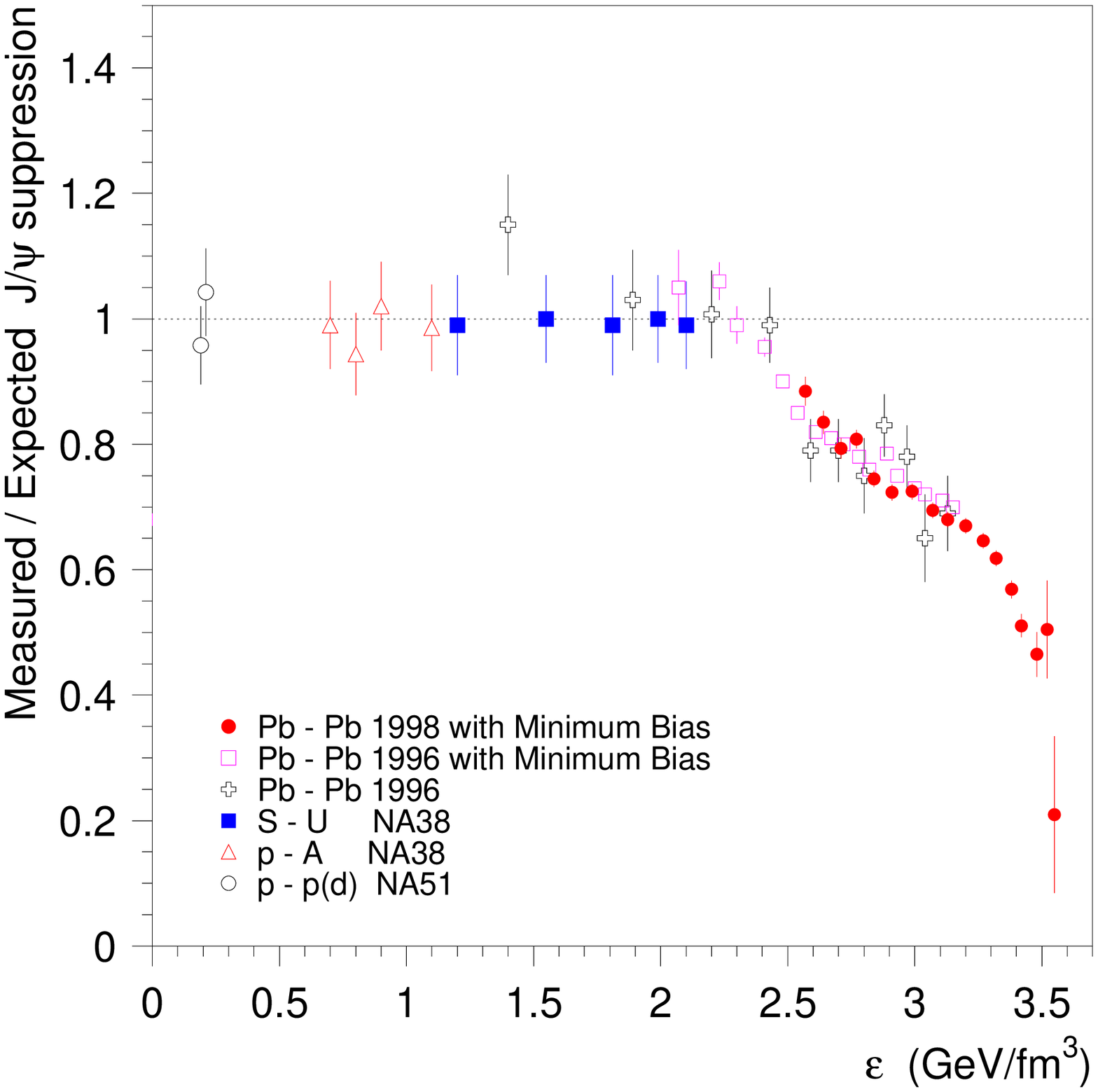}
\hspace*{-0.4cm}
  \includegraphics[height=.29\textheight]{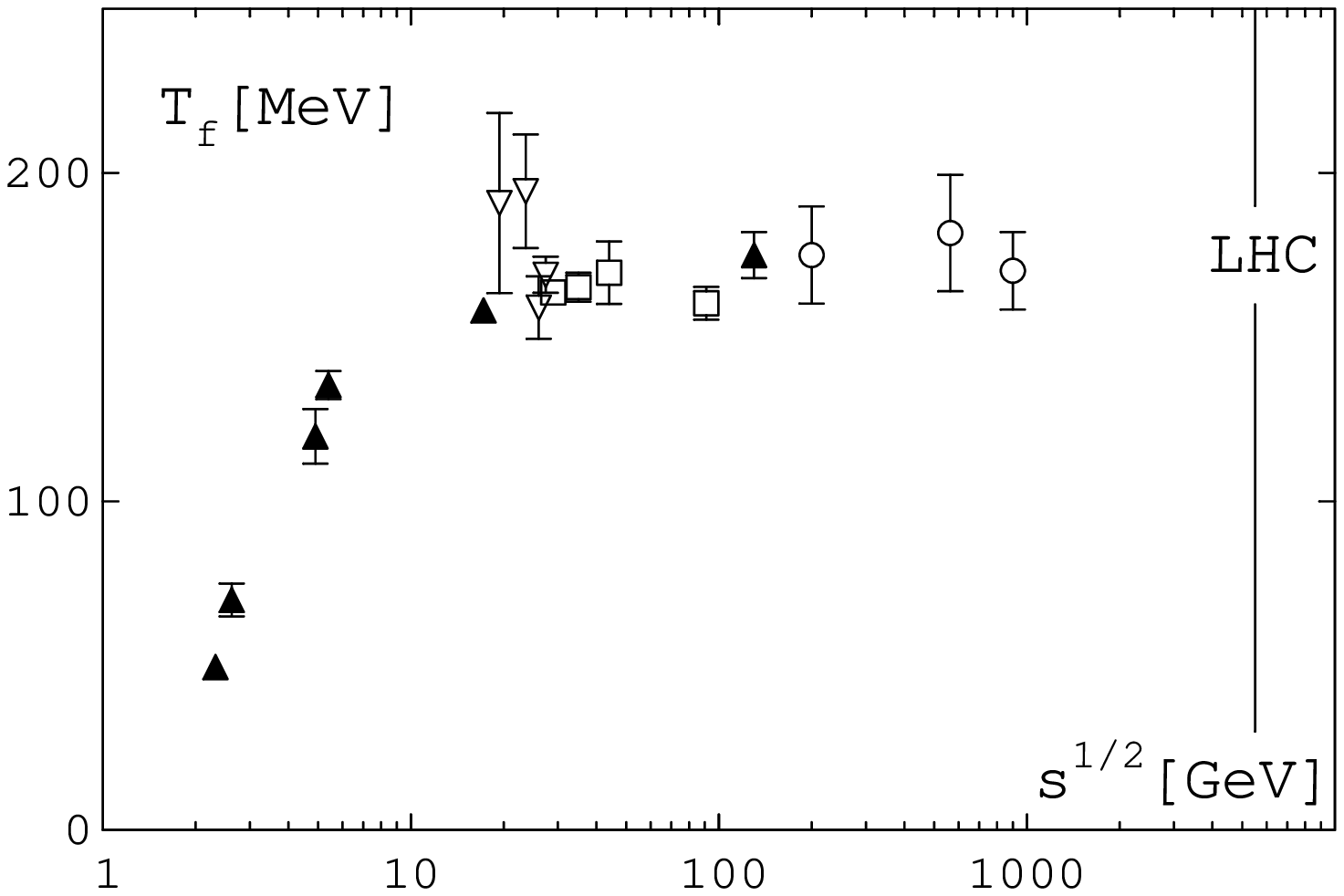}
  \caption{Left: $J/\Psi$ suppression as a function of the initial energy density (NA50 Coll.)
Right: The freeze out T as a function of $\sqrt{s}$.}
\label{satz}
\end{figure}

\noindent
The phase transition between deconfined quarks and gluons and 
 hadrons at a temperature of approximately 150-200 MeV is a fundamental  prediction of QCD 
\cite{lattice}.
This phase transition is believed to have taken place in the early universe 10$^{-6}$ sec after
the Big Bang.
The transition temperature is expected to be reachable with today's accelerators
and experimental investigations have tried to induce this transition
colliding the largest available nuclei like Pb and Au at ultrarelativistic energies.
\noindent
The experimental searches for the QCD phase transition
 started  in the eighties in 
 AGS, SPS and continue at RHIC
while in the future a large experimental program is approved to continue these searches
at the LHC and the future GSI accelerator.
\\
Signatures  of this phase transition have been measured first at SPS
(2000)  \cite{sps} and then at RHIC (2003) \cite{jet_quenching}.
Among
 those signatures we remark the suppression of $J/ \Psi$ in Pb+Pb
collisions above initial energy density $\epsilon_{i}$ =2.2 GeV/$fm^3$
\cite{na50}
(fig. \ref{satz} left)
as predicted by \cite{satz},
and the rise and saturation of the chemical freeze-out Temperature
(fig. \ref{satz} right, from \cite{satz_fig})
as a function of collision energy as predicted by \cite{van_hove}.
The expectation of a limiting freeze out
 Temperature is a direct consequence of the assumption
of a phase transition in which case $T$(hadrons) is limited by  $T_{crit}$,
and would be expected  independent of the concept of  Hagedorn's limiting Temperature
for a non interacting system of hadrons.
\\
Hadronic reactions have been associated with an underlying thermodynamic behaviour
from the fifties \cite{fermi}.
Remarkably,
  A+A systems at several energies have been found to agree with a grandcanonical
ensemble,
allowing to define a temperature,
and even more remarkably
the temperatures at chemical freeze out at $\sqrt{s}$  e.g.  17 GeV
were found to be $\sim$ 170 MeV and therefore near $T_{crit}$
 \cite{pbm_first_thermal,mapping,thermal}.
 Deviations  seen e.g. in resonances are under investigation and can be due e.g. to
rescattering of their decay products in the source   \cite{resonances}.
Certain colliding systems e.g. the ones taken with minimum bias triggers, 
do not agree with a grandcanonical ensemble.
While we discuss these deviations very briefly,
we mainly focus here on strangeness production in
 colliding systems which agree with a grandcanonical ensemble.
\\
 
\begin{figure}
\includegraphics[height=.28\textheight]{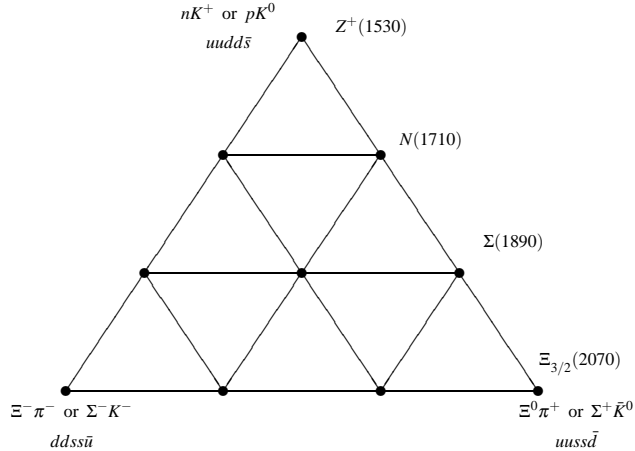}
  \caption{Predicted pentaquark antidecuplet with spin-parity $1/2^+$.}
\label{anti-10}
\end{figure}

\noindent
Pentaquarks is a name devoted to describe baryons made by 4 quarks and one antiquark e.g. 
$uudd \overline{s}$.
Exotic pentaquarks have an antiquark with different flavour than their quarks.
For recent reviews on pentaquarks see \cite{jaffe_review,pdg}.
Pentaquarks
have been predicted long time ago e.g. \cite{pq_prediction,pra} while  the number of pentaquark multiplets and their characteristics vary
 depending
on the model used.
Figure \ref{anti-10} shows the prediction of 
a spin-parity $1/2^+$ antidecuplet  of pentaquarks 
within the chiral soliton model
\cite{diakonov_polyakov_petrov_1997}.
\noindent
Pentaquark searches were performed already in the 60'ies but few low statistics candidates
found  have not been confirmed \cite{pdg_1986}.
However, recent significant advances in theoretical \cite{diakonov_polyakov_petrov_1997}
 and experimental work
 led to
 a number of new  candidates in the last 2 years of searches 
\cite{ nakano, clas_1, saphir,  clas_2, diana, hermes, neutrino, zeus, cosytof, camilleri,
 na49, na49_ximinus, h1, jamaica}.
\\

\noindent
Both the relevance of strangeness as
 evidence
for the QCD phase transition
as well as the evidence for the existence of pentaquarks are 
 unsettled problems. 
Their clarification will improve our  understanding of non perturbative QCD
in an important way.
QCD, while well understood at high energy scales and small couplings,
is less comprehensible at low energy scales and large couplings.
Basic questions like what makes up most of the nucleon mass, beyond the small part
originating from the Higgs mechanism, search their answer in understanding
non perturbative QCD.
 The confirmation of pentaquarks would break open the 
over many decades prevailing picture of hadrons as $qqq$ and $q \overline{q}$ constructions only,
while another fundamental prediction of QCD namely the existence of glueballs,
still strives to be definitively answered.
The slow and difficult progress on all the above questions over the years,
reflects the fact that the
 discovery and understanding of both the QCD phase transition as well as
 of certain rare exotic hadronic states like pentaquarks, hybrids and
glueballs  represent an enormous experimental and theoretical challenge.
\\

\noindent
In section 2 we will discuss the initial prediction of
 strangeness enhancement as QGP signature (2.1),
and review the observations, their interpretations
and point to some open questions (2.2, 2.3, 2.4).
We address especially the collision energy and energy density dependence
of the total $s \overline{s}$ production
as compared to light $q \overline{q}$ pairs
in colliding systems which agree with a grandcanonical ensemble (2.2).
This is then linked to the energy and energy density dependence
of the chemical freeze out temperature and ultimately,
to the QCD phase transition (2.3).
In (2.4) we address briefly some selected topics concerning
strangeness in QGP, reflecting work in progress.
\\
In section 3, we review the experimental data on pentaquark candidates,
as well as non-observations.
We discuss to which extend these sometimes contradicting 
informations may   lead to a consistent picture.

\section{Strangeness and the QCD phase transition}

\subsection{Prediction: the strangeness enhancement}

\noindent
Hadrons with strange quarks have been predicted to be produced at an enhanced rate
out of a Quark Gluon Plasma \cite{rafelski}.
$S \overline{s}$ pairs can be produced in the QGP 
through gluon fusion with a low threshold defined by the mass of the s quark
which due to chiral symmetry restoration is decreased to $\sim 150 $ MeV.
The mass of the s quark is similar to the transition temperature 
$m_s \sim T_{crit} \sim 150 MeV $ and  strangeness production is expected to reach equilibrium.
Multistrange baryons and antibaryons  
are expected to be even more enhanced out of a QGP because due to their heavy
mass they are hardly expected to reach equilibrium in a system below $T_{crit}$.
Therefore the strangeness enhancement is expected to rise with the strangeness content.
We will present and  discuss in the following selected experimental
results  focusing on two particular questions: Firstly, if strangeness is enhanced 
and as compared to what and
secondly, if  there is experimental evidence for the QCD phase transition from
strangeness production.

\subsection{Is strangeness enhanced and as compared to what?}

\noindent
{\bf The first measurements and first definition of strangeness enhancement}

\noindent
Historically, strange particle production has been measured in several
collisions of heavy nuclei (A+A) 
and has been systematically
 compared to p+p, p+A  or peripheral A+A collisions at the same energy.
The collisions of elementary particles like p+p or of peripheral
A+A collisions have been taken in the literature as a reference system, that is 
they have been  assumed to represent a colliding system in which no phase transition
takes place because the volume is too small.
In this case the p+p system is not expected to reach equilibrium.
\\

\noindent
The
 first observation of a $K/\pi$ enhancement was reported in 1988  in central
Si+Au collisions at $E_{beam}$ 14.6 A GeV
 \cite{e814}.
The first observation of a  $\Lambda$, $\overline{\Lambda}$ and $K^0_s$
over pion
 enhancement in A+A was reported in 1991 in central S+S collisions
at $E_{beam}$ 200 A GeV, 
compared to peripheral S+S and p+S collisions 
 \cite{na35_ss}. These first results were followed by 
a vast number of other observations of strangeness enhancement, e.g. \cite{rest}.
\\
Therefore the 'strangeness enhancement' has been first observed and defined as
an  enhancement of the double ratio of strange particles over pions
in central
 A+A collisions as compared to p+p, p+A or peripheral A+A collisions at the same energy.

\vspace{0.4cm}
\noindent
{\bf Strange baryons and antibaryons}

\begin{figure}
 \includegraphics[height=.28\textheight]{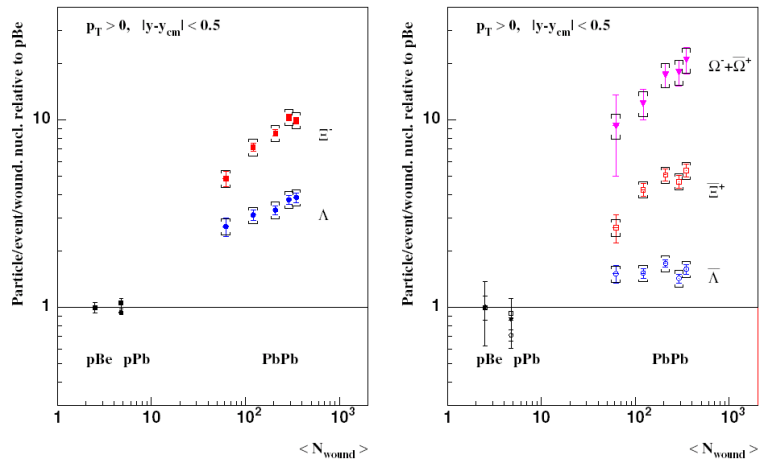}
  \caption{Particle yields 
in Pb+Pb collisions at $\sqrt{s}$=17 GeV from the NA57 experiment
shown per participant nucleon N,
as a function of N.
The ratios have been normalized to the same ratios in p+Be collisions at the same energy
also from NA57.
}
\label{xi_158}
\end{figure}

\noindent
 A  spectacular later finding 
was the enhancement factors up to 20 which have been measured
for multistrange (anti)baryons ($\Xi$, $\overline{\Xi}$, $\Omega$, $\overline{\Omega} $)  in Pb+Pb collisions at $\sqrt{s}$= 17 GeV
as compared to p+A collisions at the same energy
(figure \ref{xi_158})
 \cite{bruno_qm2004}.
While $\Xi$ production at AGS \cite{xiags} is 
well described with hadronic models this is not the case at SPS
energy  \cite{antinori}.

\begin{figure}
  \includegraphics[height=.28\textheight]{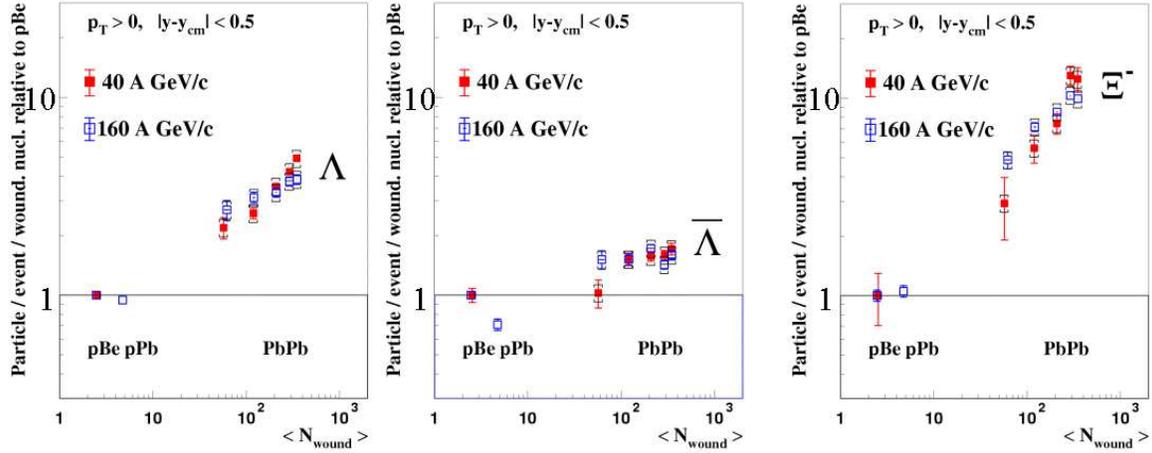}
  \caption{Particle yields per participant nucleon N, as a function of N
in Pb+Pb collisions at $\sqrt{s}$=17 and 8.8 GeV from the NA57 experiment.
The ratios have been normalized to the same ratios in p+Be collisions at the same energy
also from NA57.
}
\label{xi_40_158}
\end{figure}

\begin{figure}
  \includegraphics[height=.28\textheight]{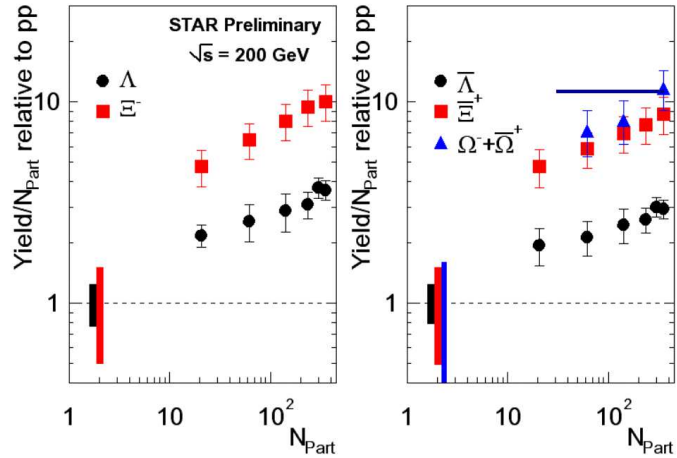}
  \caption{Particle yields 
in Au+Au collisions at $\sqrt{s}$=200 GeV from the STAR experiment
shown per participant nucleon N,
as a function of N.
The ratios have been normalized to the same ratios in p+p collisions at $\sqrt{s}$=200 GeV
also from STAR.
}
\label{star_n}
\end{figure}

\noindent
It is seen that the enhancement rises with the strangeness content.
The 'strangeness enhancement' has been defined here as ratio of
strange particle  to the number of participant nucleon (N) in Pb+Pb
as compared to p+A collisions at the same energy.

\noindent
Another remarkable observation is that the yield of $\Xi$ and $\Lambda$
  remains constant in $\sqrt{s}$= 8.8, 17 and 200 GeV 
\cite{elia_qm2004}.
This may  originate from a connection of strange baryons to the initial baryon number.
It may however also 
be an artifact of the phase space acceptance of the measurement  and  the different baryon
distributions at SPS and RHIC,   as
many strange baryons are expected where the baryons are abundant e.g. 
at forward rapidities. This can be studied with $\Lambda$'s in the FTPC of STAR
and  BRAHMS forward $K$ data  \cite{brahms}.

\vspace{0.4cm}
\noindent
{\bf N dependence}

\noindent
Baryons as $\Xi$, $\Lambda$ and $\Omega$ (the sum $\Omega$+$\overline{ \Omega}$ at this energy represents mainly
the $\Omega$'s) per participant N in Pb+Pb at 158 A GeV
  rise with N, while antibaryons ($\overline{ \Xi}$, $\overline{ \Lambda}$)
 show a saturation in the most central bins (fig. \ref{xi_158}).
This
 saturation maybe due e.g. to antibaryon annihilation in the highest densities.
Since particle yields per N is approximately proportional
to number density, assuming that  N is proportional to the Volume
(which may be incorrect due to expansion)
in equilibrium particles per N should be N indepandant.
Therefore a rising particle per N ratio versus N could be interpreted 
as a deviation from equilibrium \cite{charm}.

\noindent
The
 rise and saturation of particle densities with increasing volume as
seen in the antibaryons, 
has been described
 as due  to the gradual change from a canonical to a grandcanonical equilibrium
description e.g. \cite{redlich}\footnote{In this work, with "canonical"
describtion it is meant that the system did not yet reached the thermodynamic
limit at which all ensembles are equivalent.} 
However, the latter two cases  should be visible in the baryons too, not only in antibaryons
which is not the case.

\noindent
Another possibility is that the change of density with N is due to a rise of
temperature \cite{charm} and not to deviations from full equilibrium,
 assuming that an equilibrium description is valid for each of those N points to be able
to define T.

\noindent
The rise of particles per N with N may be attributed to 
 hard collisions.
However, the rise of particles per N with N seems steeper at 40 GeV than at 
170 GeV Pb+Pb collisions (fig. \ref{xi_40_158}).
If the deviation from flat behaviour was due to a hard component, it should
increase with energy.
Therefore the rise seen in 40 and 170 GeV PbPB is not
due to a hard component in strangeness production at these energies.

\noindent
At RHIC (figure \ref{star_n})
one observes a continous rise of particles per N  \cite{helen}.
The $\overline{\Lambda}$ do not saturate, neither the  $\overline{\Xi}$.
At RHIC one may expect hard processes to manifest themselves
and this can be investigated looking at the $p_T$ spectra and
the  $ N^{ \alpha}$ dependence of particles 
and comparing to lower energies and  to models.

\noindent
In
 conclusion the strange baryons per N rise with N at all energies ($\sqrt{s}$= 8.8, 17 and 200 GeV).
Antibaryons do  so only at RHIC, while at lower energies show saturation.
The latter can be understood as due to antibaryon annihilation.
The understanding of the N dependence of strange and non-strange particles
is an open issue which needs more data and theoretical work to be understood.

\begin{figure}
  \includegraphics[height=.32\textheight]{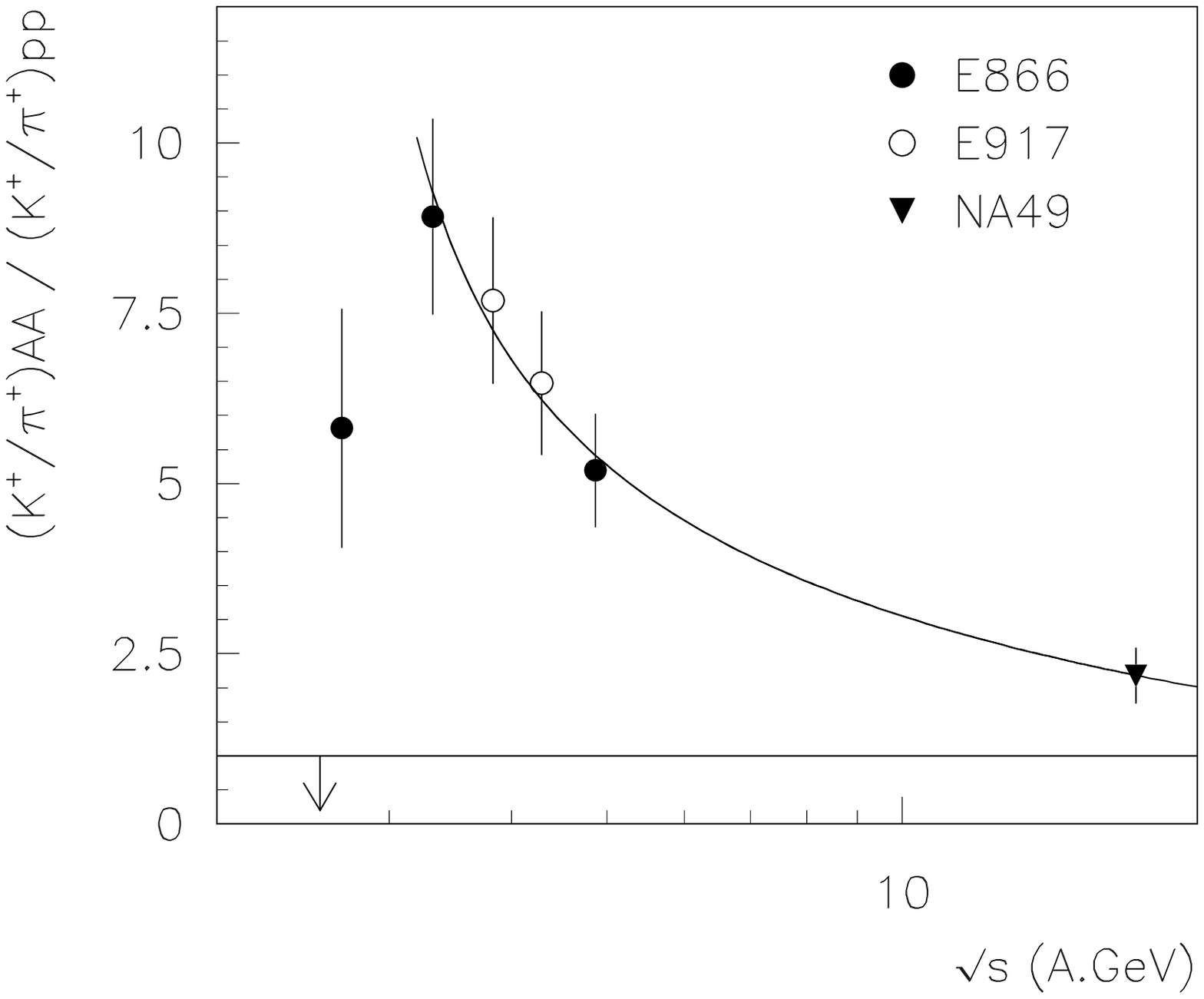}
  \includegraphics[height=.32\textheight]{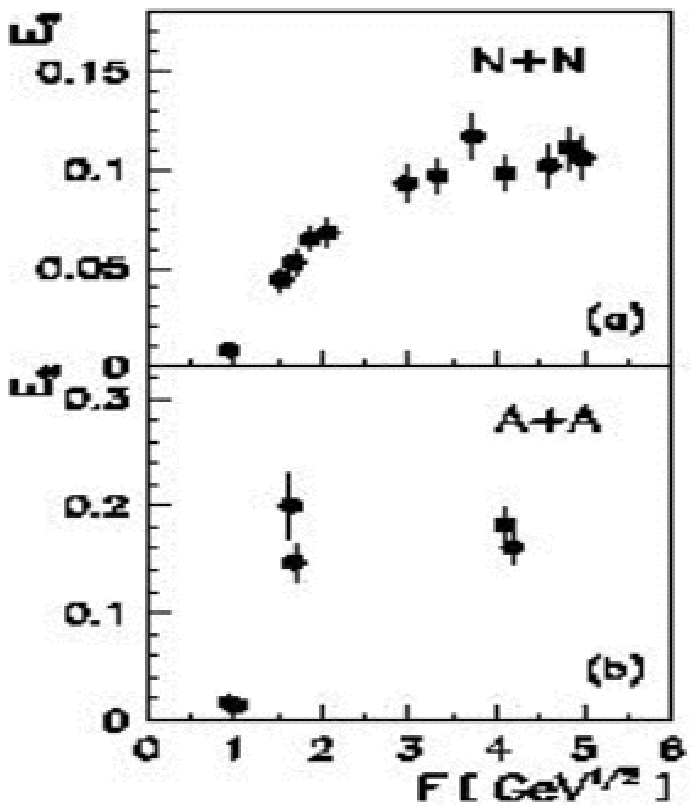}
  \caption{Left: Double ratio of kaon to pion in A+A over p+p collisions as a function
of energy. 
Right: $E_s = (\Lambda + K) / \pi$ ratio as a function of energy (F=f($\sqrt{s}$).}.
\label{ogilvie}
\end{figure}

\vspace{0.4cm}
 \noindent {\bf Is the observed enhancement as expected for the QCD phase transition ?}

\noindent
The enhancement factors of strange baryons and antibaryons 
  are slightly smaller at $\sqrt{s}$= 17 GeV than at 8.8 GeV, and
are
smaller at $\sqrt{s}$= 200 GeV (STAR) than at 17 GeV (SPS),
 with the exception of $\overline{\Lambda}$.
This exception  maybe due to annihilation effects
and the different $\mu_b$ leading to a larger $\overline{ \Lambda} / \Lambda $  ratio at RHIC.
We 
therefore observe, that the enhancement factors of strange baryons and antibaryons per N in Pb+Pb 
as compared to p+A at same energy seem to decrease with increasing energy.

\noindent
Figure \ref{ogilvie} left 
 shows that  the K/pi ratio in A+A over p+p collisions
\cite{ogilvie}
is as well increasing towards lower energies.
Why is that so?
We will address this question at the end of section 2.2.
\\

\noindent
Remarkably, 
this behaviour is exactly opposite to the theoretical expectation that $s \overline{s}$ enhancement is
induced with increasing T and collisions energy, namely at or near Tc.
\\
This disqualifies  the strangeness enhancement definition as enhancement in central A+A collisions
over p+p or p+A collisions at the same energy, as a QGP signature.

\vspace{0.4cm}
 \noindent {\bf Energy dependence of  strange to pion ratio in A+A collisions}

\noindent
Another approach to study strangeness production relative to light flavours (u,d) is
to look at the energy dependence of strange to non strange particles in the same
kind of reactions e.g. nucleus+nucleus and particle collisions separately.
The first  such study \cite{dieter}  has shown
that the energy dependence of strange particle to pion ratio 
($E_s$) is different in  A+A and N+N collisions.
They observed a rise and subsequent saturation of $E_s$ with energy (fig. \ref{ogilvie}, right).
The
 authors proposed that the increase of strangeness over pion ratio
at low energies
is a consequence of the strangeness over entropy ratio being proportional to
the chemical freeze-out temperature, which rises with collision energy.
They observe a maximum of $E_s$ in A+A, unlike p+p collisions, 
and 
they interpret this as due to the onset of the phase transition taking place
between AGS and SPS near the maximum of $E_s$ \cite{marek_gorenstein} which had yet to be localized
at that time.

\begin{figure}
  \includegraphics[height=.25\textheight]{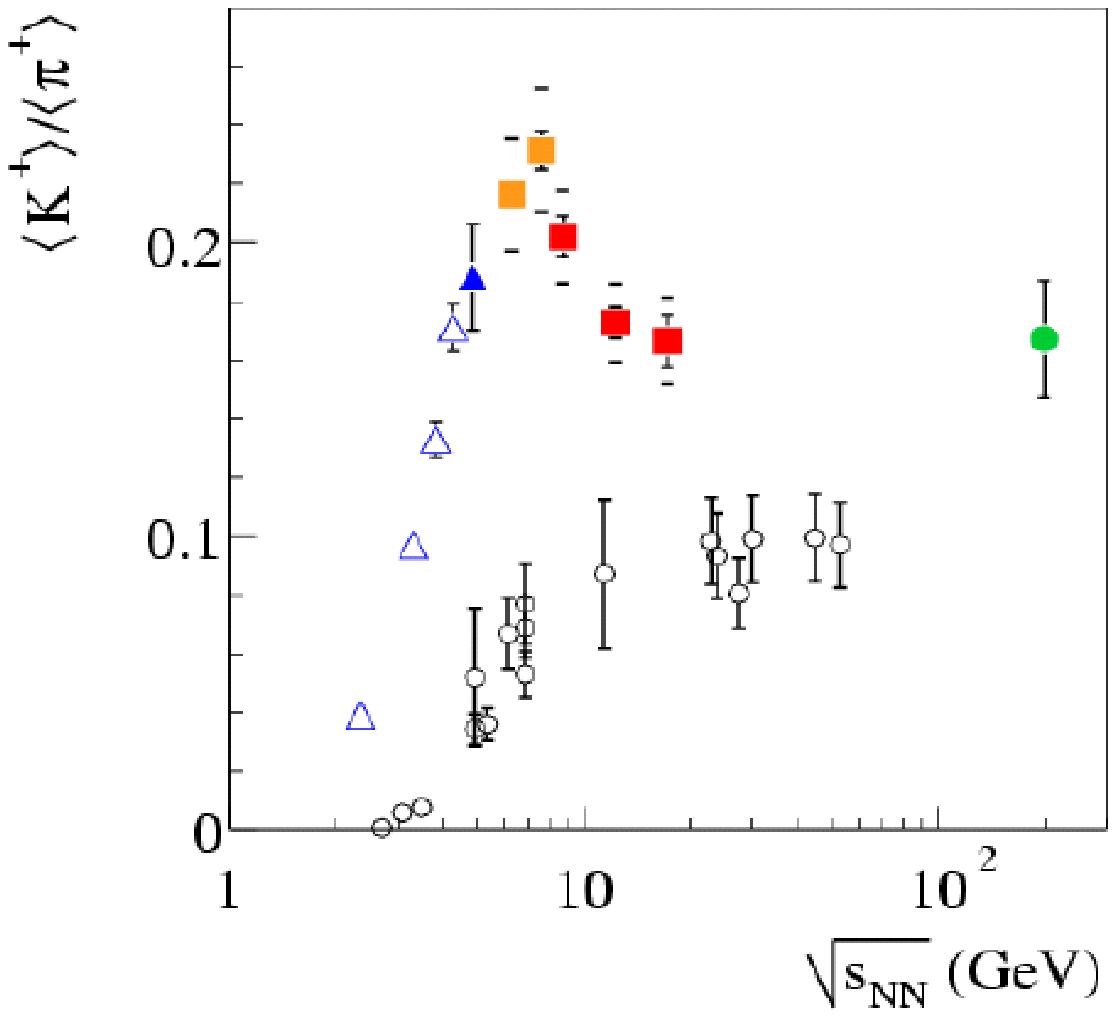}
\hspace*{0.6cm}
  \includegraphics[height=.25\textheight]{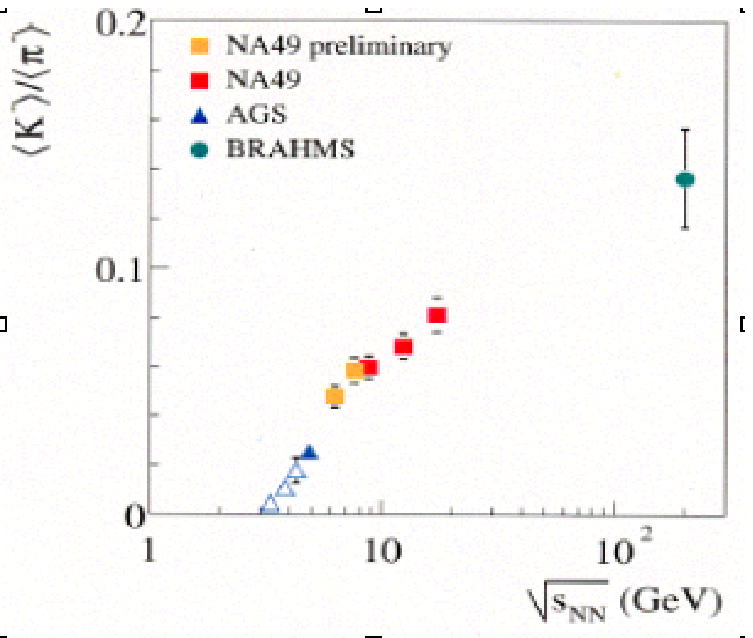}
  \caption{Left: $K^+/\pi^+$ ratio as a function of $\sqrt{s}$ in several reactions. 
Right:  $K^-/\pi^-$ ratio as a function of $\sqrt{s}$ in several reactions.}
\label{na49_1}
\end{figure}

\begin{figure}
  \includegraphics[height=.25\textheight]{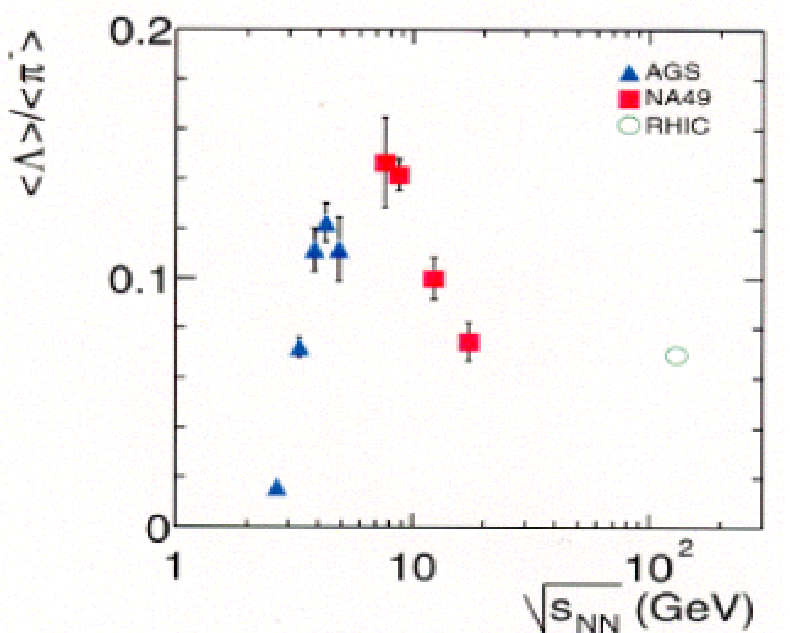}
\hspace*{0.5cm}
  \includegraphics[height=.25\textheight]{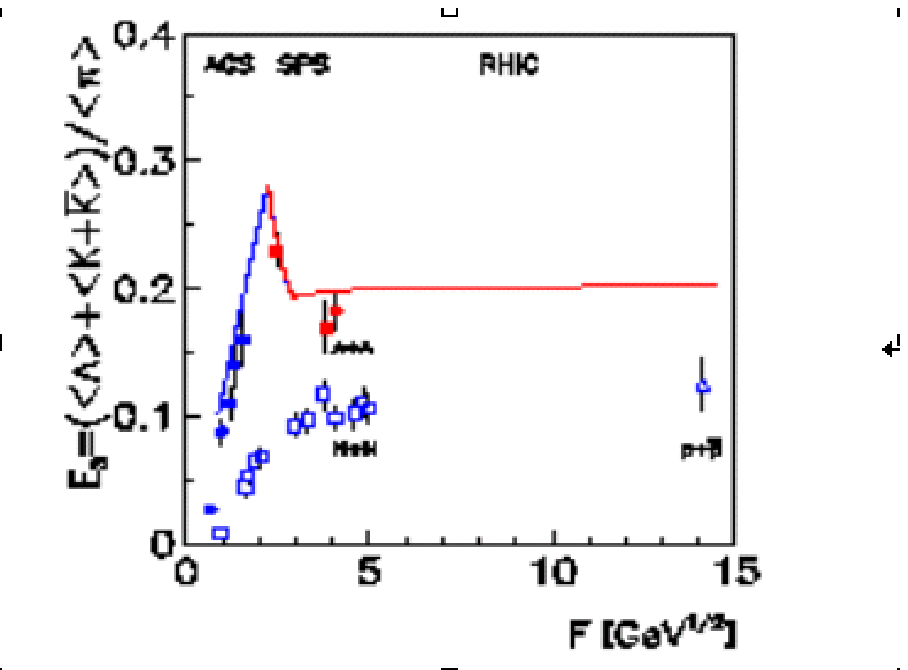}
  \caption{Left: $\Lambda /\pi$ ratio as a function of $\sqrt{s}$ in several reactions. 
Right:  $E_s$ factor as a function of $\sqrt{s}$ in several reactions. See text for more explanations.}
\label{na49}
\end{figure}

\vspace{0.4cm}
 \noindent {\bf Maximum of strange/pion ratio at 30 GeV Pb+Pb collisions}

\noindent
In the mean time, the maximum seen in $E_s$ has been confirmed dramatically through 
latest data from NA49 \cite{na49_qm2004}. NA49 
found a maximum at 30 GeV Pb+Pb collisions (figure \ref{na49_1}, \ref{na49})
however only in $K^+/\pi^+$, $\Lambda/\pi$, while no maximum is seen in
$K^-/\pi^-$ which rises continouisly with energy (figure \ref{na49_1} right).
Since a high baryon to antibaryon ratio favours the associated production of $\Lambda K^+$
the above differences
 suggest that they may be connected to the different baryochemical potentials $\mu_B$
reached at different energies in the same collision system (Pb+Pb, Au+Au) near midrapidity.

\noindent
The
 baryon to antibaryon ratio is continouisly
droping with decreasing energy in the same A+A system (e.g. \cite{elia_qm2004}).
 Which implies that the $K^+/\pi^+$ and  $\Lambda/\pi$  ratio
should continouisly rise with decreasing energy. Why is this not the case?
Because the temperature drops below a certain energy (figure \ref{satz}, right)
and all ratios reflect this fact.

\noindent
The hadronic models URQMD
and HSD underestimate the $K^+/\pi^+$ ratio
but do reproduce well the $\Lambda/\pi$ ratio and it's maximum \cite{horst_this_conf}.
To understand this discrepancy one could investigate the sharing 
of $s \overline{s}$ among hadrons,  check conservation of  $s \overline{s}$
 and study the dependence of
 major channels like $K^+ K^- $ and  $ \Lambda K^+$
 from collision  energy, stopping and baryon density.
For  more discussion of hadronic models see e.g. \cite{antinori,horst_this_conf}.
\\

\noindent
The
 maximum of the strange to pi ratio near 30 A GeV Pb+Pb collisions
as mentioned previously
has been interpreted  as due to the onset of the QCD phase transition.
Figure \ref{na49}, right shows the prediction of \cite{marek_gorenstein}.
\\

\noindent
However, this maximum has been explained in the mean time
 as a consequence of different $\mu_B$ values.
\\
The influence of a varying $\mu_B$ in the 
energy dependence of strange to pion ratios 
has been investigated for the first time in 
\cite{mapping,border}.
Each of the studied thermodynamic systems has been extrapolated to 
the equivalent thermodynamic systems at zero chemical potential.
This is equivalent to an extrapolation of
 e.g. all Pb+Pb collisions to $Pb + \overline{Pb}$ collisions.

\begin{figure}
  \includegraphics[height=.28\textheight]{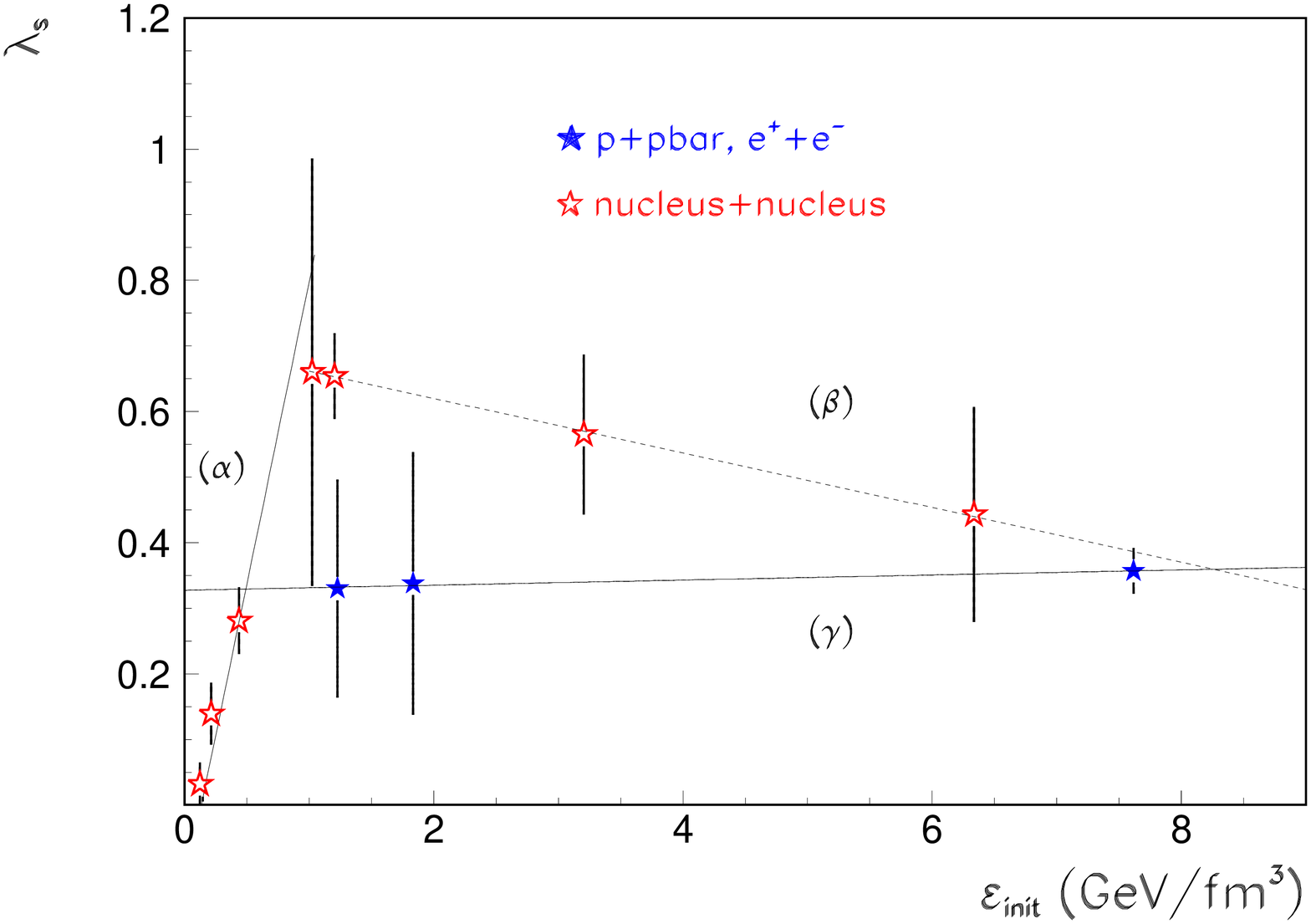}
  \includegraphics[height=.28\textheight]{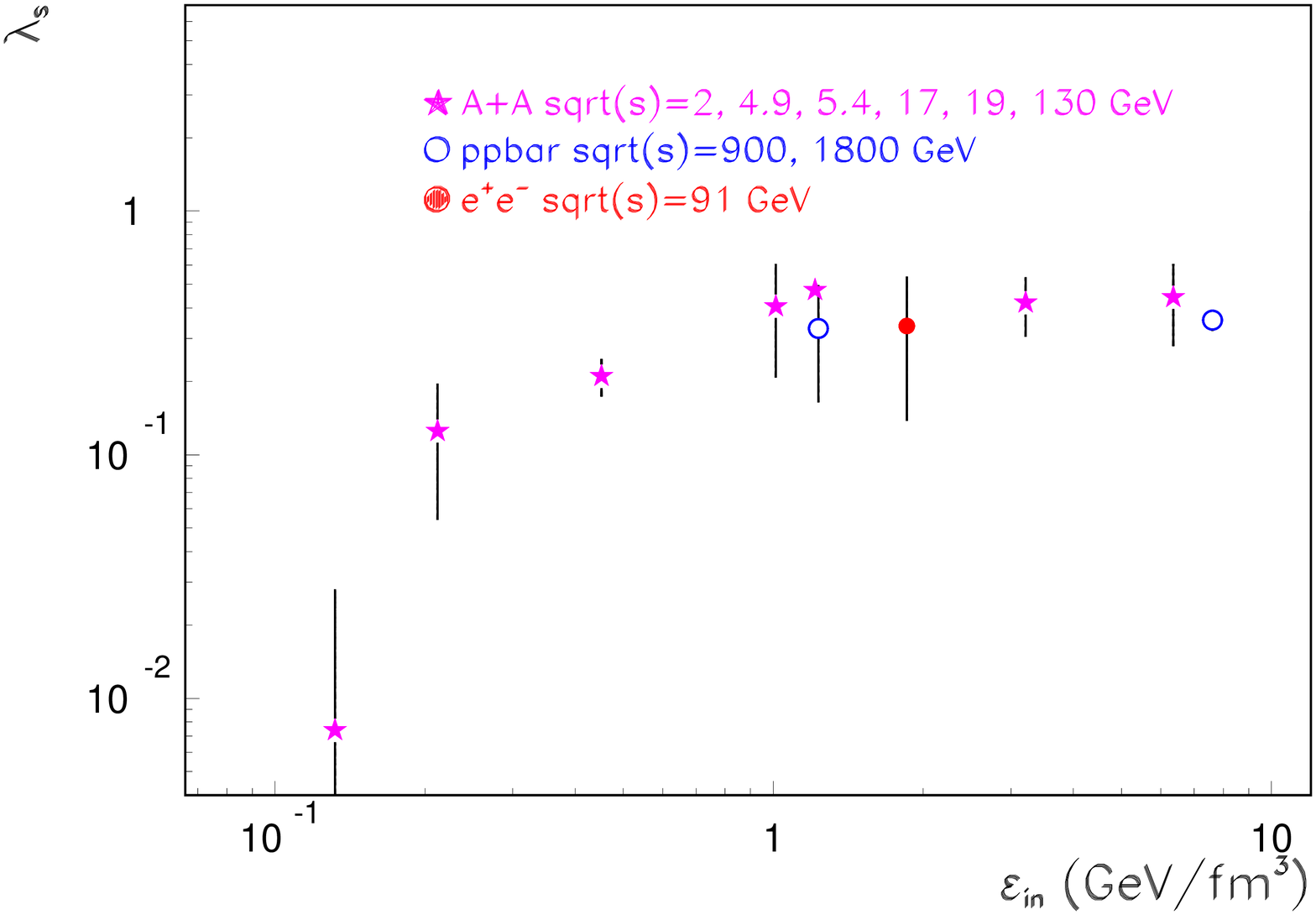}
  \caption{Left: Strangeness suppression factor $\lambda_s$ as a function of the
initial energy density in several reactions.
Right: the same as left, after extrapolating all points to $\mu_B=0$.}
\label{ls_1}
\end{figure}

\noindent
The strangeness suppression factor $\lambda_s$, defined as the ratio of newly produced
$s \overline{s}$ to light newly produced
 $q \overline{q}$ 
($\lambda_s = \frac {2 s \overline{s} } { (u \overline{u} + d \overline{d}) }$) \cite{wrow},
 is maximal in Pb+Pb collisions at 40 A GeV (figure \ref{ls_1}, left, point
at the maximum of the triangle).
However, after the difference in the $\mu_B$ of the two systems is eliminated
$\lambda_s$  is found to be the same at 40 and 158 GeV (figure \ref{ls_1}, right).
Therefore the  difference in strangeness 
is explained as due to the higher $\mu_B$ in 40 GeV as compared to the 170 A GeV Pb+Pb collisions.
\\

\noindent
If a part of the $K/\pi$ enhancement is due to the onset of the
 QCD phase transition and a part to the high baryon density,
 the extrapolation to zero potential of all systems \cite{mapping,border}
should leave a residual enhancement due to the phase transition itself.
Figure \ref{ls_1}, right, shows no such enhancement.
This means that the enhancement seen in the $K/\pi$ 
and $\Lambda/\pi$ ratios at 40 GeV Pb+Pb collisions
(figure \ref{na49}, \ref{na49_1})
has nothing to do with the phase transition. 
One should look if a peak appears after eliminating the $\mu_b$ 
 with more data at 20 and 30 GeV (SPS) and around that energy at the future GSI.
This has not been studied yet.
The experimental measurement of many particle ratios at 20 and 30 A GeV Pb+Pb
collisions by NA49 is therefore important for the understanding of this maximum.
\\

\begin{figure}
  \includegraphics[height=.30\textheight]{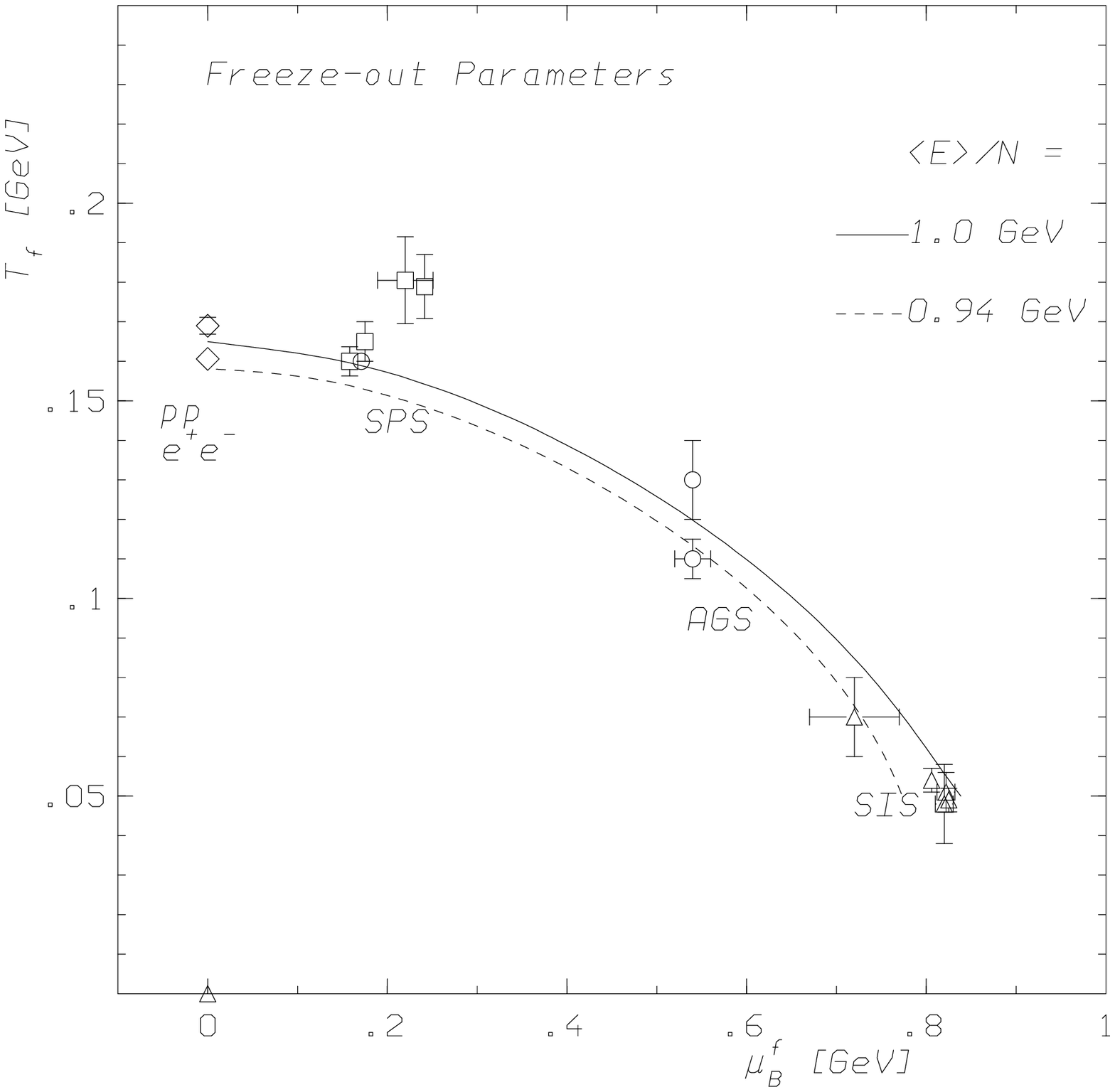}
\hspace*{0.6cm}
  \includegraphics[height=.30\textheight]{figures.d/pbm_max.eps}
  \caption{Left: T versus $\mu_B$. 
Right: $\lambda_s$ versus energy. 
See text for explanations.
}
\label{ls_2}
\end{figure}

\noindent
The 
same conclusion that the 'maximum' of strange to pion ratio is driven by the
baryons  has been reached later by \cite{pbm_max}.
The (T,$\mu_B$) points describing colliding systems
were found interestingly to be fitted with a curve assuming constant energy per hadron 
at freeze out  $<E>/N$ $\sim$ 1 GeV \cite{cleymans} (fig. \ref{ls_2}, left).

\noindent
Figure \ref{ls_2}, right,
shows that this parametrization,  while it describes  the general
trend of the  strangeness suppression factor
versus energy, it does not fit in detail the $\lambda_s$ e.g. at SPS.
In addition there is no pronounced peak but a broad maximum.
\\

\noindent
However, one should not conclude from this, that thermal models do not reproduce the maximum.
The lack of a pronounced maximum is  not due to the thermal model
since
the 30 and 40 A GeV points are well described by thermal model fits
\cite{manninen,border,becatini_40}.
The less accurate description of the energy dependence of  $\lambda_s$
and the lack of a sharp maximum,
may be due to the fact that the parametrization of a constant $<E>/N$ $\sim$ 1 GeV
does not describe well the ratio of strange to light quarks $\lambda_s$.
Note that the ratio  $<E>/N$ is not a constant but exhibits a change in other thermal
model analysis (fig. 14 of
reference  \cite{border}).
\\

\begin{figure}
  \includegraphics[height=.29\textheight]{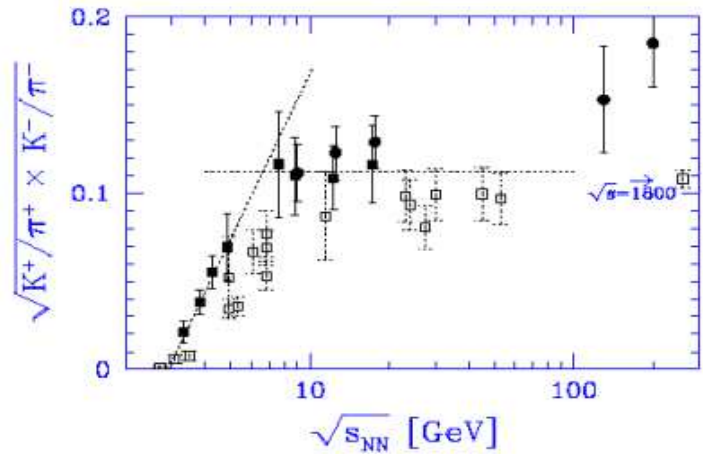}
  \caption{The quantity $\sqrt{ ( K^+/ \pi^+ ) (  K^-/\pi^-) }$ as a function of energy.
}
\label{ls_3_rafelski}
\end{figure}

\begin{figure}
  \includegraphics[height=.4\textheight]{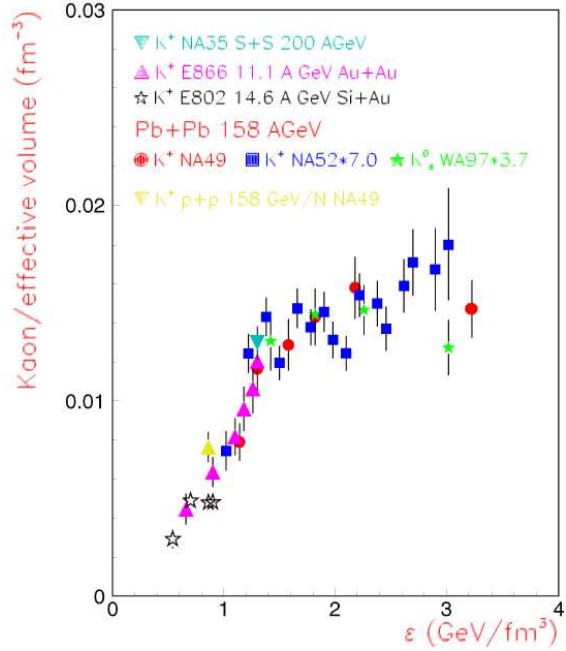}
  \caption{
Ratio of kaon yields per effective Volume as a function of the
initial energy density for several reactions.}
\label{ls_3}
\end{figure}

\noindent
In another study, the
 ratio $\sqrt{  (K^+/ \pi^+ ) (  K^-/\pi^-)  }$ is found to rise and saturate with increasing energy
while increasing at RHIC energy \cite{rafelski_k} (figure \ref{ls_3_rafelski}).
This ratio is special, as it constructed in such a way that it is  $\mu_B$ independant.
The lack of any maximum near 40 GeV Pb+Pb strengthens the previous conclusion
that the maximum is due to the different $\mu_B$ \cite{border,pbm_max}.
\\

\noindent
In \cite{rafelski_k} the higher $K/\pi$ at RHIC is discussed as possibly reflecting 
a  'strangeness enhancement', seen only in Au+Au collisions at RHIC as compared to
$ p+ \overline{p}$ collisions (fig. \ref{ls_3_rafelski}). 
There are two things to mention.
Firstly,  since the share of $s \overline{s}$ among hadrons 
changes with energy  one should look at the total strangeness
production ($\lambda_s$) versus energy. 
\\
When one does this, looking at the $\lambda_s$ at $\mu_B=0$
 (fig. \ref{ls_1}, right) this enhancement dissappears.

\vspace{0.4cm}
 \noindent {\bf Strangeness in $ p \overline{p}$ collisions at the Tevatron is similar to Au+Au at RHIC?}

\noindent
Secondly,
the point for $p \overline{p}$ collisions at
$\sqrt{s}$ = 1.8 TeV (last open rectangular point to the right)
in figure \ref{ls_3_rafelski}
is for a minimum bias trigger.
The minimum bias point is below the points for central Au+Au collisions at RHIC.

\noindent
However,    for $p \overline{p}$ collisions at $\sqrt{s}$ = 1.8 TeV
with the highest charged multiplicity,
one finds that
the resulting $K/\pi$ ratio of 0.14 $\pm$ 0.2 is consistent with the RHIC Au+Au data
shown in fig. \ref{ls_3_rafelski} \cite{mapping}.
\\
In addition the $\lambda_s$ from a  grandcanonical fit with
good $\chi^2/DOF$ to these data
is consistent with the  $\lambda_s$ at RHIC when the different $\mu_b$ are eliminated \cite{mapping}
(fig. \ref{ls_1}, right).
While the  minimum bias $p \overline{p}$ collisions at $\sqrt{s}$ = 1.8 TeV
do not agree with a grandcanonical fit.
\\
\noindent
These
 findings 
 call for more measurements of particle ratios in
$p+p$ and $p + \overline{p}$ collisions at the Tevatron and the LHC.

\vspace{0.4cm}
 \noindent {\bf Strangeness enhancement is due to a canonical to grandcanonical transition?}

\noindent
It is discussed recently in the literature
 that within statistical models the enhancement of strangeness in A+A is the
result of volume dependant 
canonical strangeness suppression in p+p,  p+A and possibly minimum bias A+A collisions.

\noindent
This is not really true, because even if the p+p and A+A   systems
were both grandcanonical, the strangeness
would be still enhanced in A+A
as compared to p+p collisions,  if e.g. $\mu_b$ and/or the initial reached T and $\epsilon$ differ.

\noindent
At the same colliding energy  p+p , p+A and  A+A
collisions (assuming we can use thermodynamic language to describe them
and we do that when  thermal model fits work well)
form  systems with different thermodynamic properties: different baryochemical potentials,
different temperatures and  different energy densities.
Comparing therefore A+A to a p+p or p+A collision at the same energy
may be like comparing apples and oranges.
This explains why the $K/ \pi$ ratio in $A+A$ over $p+p$ collisions 
increases with decreasing collision energy as seen in fig. \ref{ogilvie}, left.
\\

\noindent
Comparing the same A+A collision system at different 
energies, may also be  like comparing apples and oranges.
For example, we have seen  that the higher strangeness content in 30 and 40 GeV Pb+Pb
as compared to 158 is due to the different $\mu_B$ \cite{border,pbm_max,rafelski_k}.
This has nothing to do with a phase transitiion.
\\

\noindent
The above thoughts  lead us to formulate some  questions: 
\begin{itemize}
\item
Is there a strangeness enhancement
beyond trivial sources causing strangeness enhancement like a higher
$\mu_B$ ?
\item
Should we compare different colliding systems at the same colliding energy
or choose a more relevant common parameter or set of parameters ?
\item
Which should be this parameter ?
\item
Is there a strangeness enhancement in A+A over p+p collisions
when they are compared at the same value
of this parameter ?
\end{itemize}
We will address these questions in the following section.

\subsection{Do we see evidence for the QCD phase transition from ssbar production?}

\vspace{0.4cm}
 \noindent {\bf The choice of the 'gauge' parameter for comparing strangeness}

\noindent
A natural choice for the 'gauge' parameter would be the initially reached
temperature. We cannot measure this quantity.
The only quantity characterizing the intial state of the colliding
system which can be estimated from measurements is the initial energy density $\epsilon_i$.

\noindent
The
 first study of strangeness production 
as a function of the initial energy density has been  performed in
\cite{charm}.
It was found (figure \ref{ls_3}) that looking at  the ratio of Kaon per participating nucleons
not as a function of participant nucleons, but as a function of $\epsilon_i$
transforms the puzzle of different N dependences seen in different energies and looking incoherent,
 into
 a universal curve throughout the different collision systems 
(fig. \ref{ls_3}).

\noindent
Most importantly 
 it
rises with $\epsilon_i$ and saturates at the vicinity of $\epsilon_i$ $\sim$ 1 GeV
which corresponds  to 40 GeV Pb+Pb,
therefore showing a dramatical change near $\epsilon_e$ $\sim$ 1 GeV/$fm^3$.
This
 onset of strangeness saturation is at a lower $\epsilon_i$ 
than the onset of $J/\Psi$ suppression of 2.2 GeV/$fm^3$ (fig. \ref{satz}) which maybe due to
the fact that $J/\Psi$ suppression may be overcritical (e.g. \cite{karsch}).

\noindent
No maximum is seen in  figure   \ref{ls_3},  for the reason
that the data near $\epsilon_e$ $\sim$ 1 GeV are not taken
from 20, 30 and 40 GeV Pb+Pb collisions but are all
from 158 GeV Pb+Pb collisions.
Therefore the $\mu_B $ does not change dramatically in the above figure.

\noindent
Note that kaons in fig. \ref{ls_3}, behave like
antibaryons in fig. \ref{xi_158}, right.
Mesons and antibaryons are less directly related to the initial
baryon number than baryons and their characteristics may deviate.
We also mentioned antibaryon annihilation.

\noindent
However, what is really relevant to study is  the overall $s \overline{s}$ trend.
\\

\vspace{0.4cm}
 \noindent {\bf Initial energy density dependence of $\lambda_s$}

\begin{figure}
  \includegraphics[height=.35\textheight]{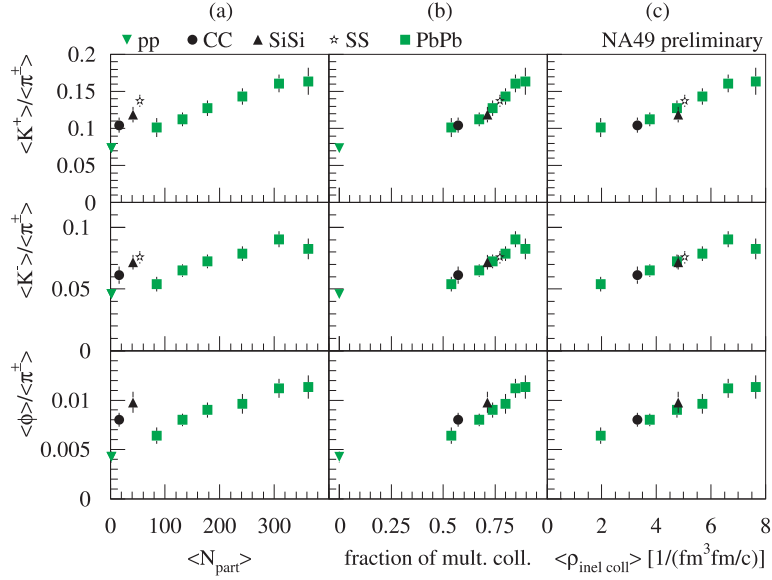}
\vspace*{-0.2cm}
\vspace*{-0.2cm}
\caption{Ratios of strange particles to pions as a function of the number
of participant nucleons, of collisions and of the space-time density of inel. collisions,
by the NA49 Collaboration.}
\label{na49_density}
\end{figure}

\noindent
The first study of the overall $s \overline{s}$ production
as compared to newly produced light quarks ($\lambda_s$) as a function of 
$\epsilon_i$ 
has been
performed in \cite{mapping,border}.
The $\lambda_s$ factor  shows a rise, a maximum 
and subsequent decrease with increasing  $\epsilon_i$
(figure \ref{ls_1}, left).
The $\lambda_s$ factor for A+A collisions is higher than
the one for p+p and $p \overline{p}$ collisions.
\\
However, after extrapolating all thermodynamic systems 
to $\mu_B$=0, the factor $\lambda_s$ shows a universal behaviour:
it rises and saturates above  $\epsilon_i \sim $ 1 GeV/$fm^3$
(fig. \ref{ls_1}, right).
This behaviour is followed by both A+A as well as by some particle
collisions like 'central'  $p \overline{p}$ collisions which give a good thermal fit.
The central Tevatron $p \overline{p}$ collisions give as mentioned the same $\lambda_s$
as central Au+Au at RHIC within errors.
A small remaining systematic
 deviation of the T and the $\lambda_s$ being somewhat smaller in $e^+ e^-$, $p \overline{p}$
and p+p collisions as compared to central A+A collisions at $\mu_B$=0
\cite{mapping},
  maybe attributed to volume effects.
\\
Practically all the enhancement of A+A over p+p collisions seen in
fig. \ref{ls_1}, left, dissappears at same $\mu_B$ (fig. \ref{ls_1}, right).
\\

\noindent
Why is the extrapolation to $\mu_B$=0 important and what we learn from that,
is discussed in the next session. 
\\
 
\noindent
The use of other scale factors like the mean space-time  density of all inelastic
 collisions during interpenetration fo the nuclei ($<\rho_{inel. coll.}>$)
 also leads to a better agreement of ratios in different A+B collisions at the same energy
(fig. \ref{na49_density}, middle and right),
which differ when studied as a function of N
(fig. \ref{na49_density}, left)
\cite{hoehne}.
However this choice of scale factors e.g. the density, does not account for the
different energy 
and therefore does not allow the comparison of collisions with different energies.
The latter
 is possible using the initial energy density, which includes the energy dependence.
\\

\vspace{0.4cm}
 \noindent {\bf The extraction of critical parameters}

\noindent
Assume we heat water, and measure the temperature versus the heat,
but  we are not allowed to measure the temperature of the steam
(analogy to QGP)
but only of the water (analogy to hadronic system).
We would measure a rising temperature and then a saturation
just below $100 $ $^0C$, at normal density and pressure for water,
 because after $T_{crit}$ is reached, 
we always measure the water temperature below $T_{crit}$
even if heating is increased.
From the plot we can  find the $T_{crit}$ for this phase transition,
as the limiting T at which the T versus heat saturates namely 100 $^0C$.
This idea has been proposed long time ago \cite{van_hove}.

\noindent
Would we now put inside the water some salt and repeat the experiment
we would each time find a different limiting T at saturation depending on
the water salinity.
Studing systems with different baryochemical
potentials is like studing systems with different salinities and searching for
the limiting temperature.
In order to measure a single limiting temperature we should
make the experiment at the same salinity.

\begin{figure}
  \includegraphics[height=.35\textheight]{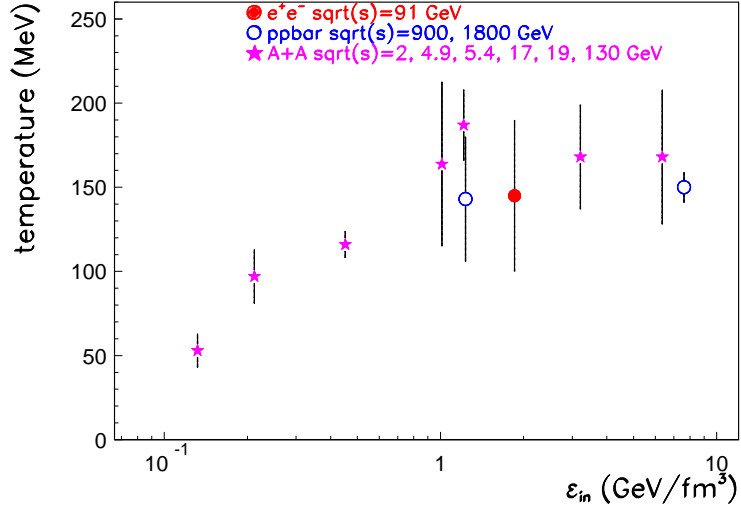}
  \caption{Left: Temperature as a function of initial energy density for several systems
at $\mu_B=0$.
}
\label{ls_4}
\end{figure}

\noindent
The simplest way to normalize nuclei and particle collisions to the same 'salinity' ($\mu_B$)
 is to take salinity  ($\mu_B$) zero.
One may then plot 
  the chemical freeze out T as a function of the initial energy density
and find the phase transition temperature
without involving a comparison to theoretical predictions.
This method has been proposed in \cite{proposal}
and was performed in \cite{mapping} and \cite{border}.
The result is as expected a rise and saturation (fig. \ref{ls_4}).
\\

\noindent
We learn (figure \ref{ls_4}) that the onset of the QCD phase transition extracted from the onset of T
saturation can be estimated to be $\sim$ 1 GeV/$fm^3$.
\cite{mapping}.
A more precise estimate leads to 0.6 $\pm$ 0.2 (stat) $\pm$ 0.3 (syst) GeV/$fm^3$ \cite{gargnano}.
This value is independent of lattice QCD predictions.
The extracted $\epsilon_{i,crit}$ is in agreement with the predicted lattice QCD values
\cite{lattice}.

\noindent
Furthermore, we  learn that the temperature does not depend on
the $\mu_B$ as sensitively as the $\lambda_s$, as the plots of T versus energy or energy density
with different $\mu_B$'s (fig. \ref{satz}, right) and with the same $\mu_B$ (=0)  (fig. \ref{ls_4}) ) are very similar.
This is why the plot of T versus collision energy at different $\mu_B$, does not show any maximum
near energy density 1 $GeV/fm^3$ (fig. \ref{satz}, right).

\noindent
On the contrary we have seen that
strangeness is very sensible to different $\mu_B$, it is a real 'baryonometer' as
well as a real 'thermometer", as it depends much also on T.

\noindent
A rising $\mu_B$ with decreasing energy would simply enhance $\lambda_s$ continuously,
(e.g. $\lambda_s$ would continue to rise also
 below $\epsilon_{i} \sim 1$ $GeV/fm^3$ in fig. \ref{ls_1}, left),
would not be the case of the T falling below a certain limit (namely exactly at 
$\epsilon_{i} \sim 1$ $GeV/fm^3$, fig. \ref{ls_4}).
\\
From the extrapolation to $\mu_B$=0 we learn that the $\lambda_s$ factor 
at  $\mu_B$=0 (fig. \ref{ls_1}, right)  shows no maximum anymore but it
simply and universally follows the T in its rise and saturation versus 
initial energy density (fig. \ref{ls_4}).
This seems to be the case for A+A and also for $p+p$ or $p\overline{p}$ 
 collisions as suggested by fig. \ref{ls_1}, while more data are needed
to study $\lambda_s$ in the latter in detail.

\vspace{0.4cm}
 \noindent {\bf Where is the strangeness enhancement and as compared to what ?}

\noindent
Strangeness enhancement
can be therefore reinterpreted as the $\lambda_s$ enhancement seen in all colliding systems
which give a good thermal fit 
and have a $\epsilon_i$ higher than approx 0.6- 1 GeV/fm3
as compared to all colliding systems which either give no thermal fit
or which have an $\epsilon_i$ smaller than 0.6- 1 GeV/fm3 (fig. \ref{ls_1}).
Therefore
 the definition of 'reference no QGP system' is 
changed in a dramatic way, as well as the notion of 'strangeness enhancement'.

\vspace{0.4cm}
\noindent
{\bf Can we measure the approach to $T_{crit}$ and the critical exponent ?}

\noindent
In addition, using parameters characterizing the data at $\mu_B=0$ one can study their
approach to $T_{crit}$ (as usually done in physics of other phase transitions)
 and extract the latter as well as the critical exponent.
The approach to $T_{crit}$ has been fitted 
with a function 
$ f= const / ln (1- T/T_{crit} )^ {\alpha}$
for the first time in \cite{border} using only data with $\epsilon_i >$  1 GeV/$fm^3$.
It was  found a $T_{crit}$= 218 $\pm$ 70 MeV and a critical exponent 
$\alpha$=0.54 $\pm$ 0.47.
The errors are very large in this study which only demonstrates the principle.
 LHC and RHIC data may allow  to study in detail the approach to $T_{crit}$
in this way.
It has been further predicted that the $\lambda_s$ factor in A+A collisions will
reach practically its limiting $\mu_B=0$ value at the LHC \cite{border}.

\vspace{0.4cm}
\noindent
{\bf How can equilibrium be 
reached in high energy particle and nuclear collisions ?}

\noindent
It is an ongoing discussion in the literature  what the saturation of the chemical freeze out T
with increasing energy really means (fig. \ref{satz}, right) and in particular
 why is the same for $e^+e^-$, p+p and A+A
collisions e.g. \cite{satz_hagedorn,dokshitzer,stock_review}.
\\
\noindent
It is  always taken for granted that no QCD phase transition
can appear in a p+p system due to the small volume.
 What if the p+p system has infinite energy ? 
After which energy the initially colliding particle volumes
plays no role anymore ?

\noindent
Maybe it is not a completely unexpected feature, if the
 hadronization of any quark and gluon system into hadrons
 (e.g. jet hadronization), 
as well as the hadronic mass spectrum, 
do reflect the existance and value of  the $T_c$ of the QCD phase transition.
Speaking about temperature, we assume implicitly thermal equilibrium.
 Could jet hadronization  be a thermodynamic process?
\\
\noindent
If a grandcanonical ensemble can describe the ratios in a $p \overline{p}$ collision
and a temperature can be defined, why not a
 "QCD phase transition" in a high multiplicity $p \overline{p}$ collision ?
\cite{gutay}
Along a hadronizing jet ?  \cite{hadronizing}.
\\
While $q \overline{q}$ produced in an elementary  collision
fly apart and cannot communicate with each other \cite{dokshitzer},
the hadronization along each jet and the resulting particle yields may exhibit 
equilibrium features reflecting the $T_c$.  
\\
Does the QCD vacuum itself have a thermostat-like nature \cite{dokshitzer} ?
\\

\noindent
However, fact is that like A+A collisions, also elementary collisions do not always
agree with a thermal model.
In particular $p \overline{p}$ collisions, like A+A collisions too,
 need a "centrality" trigger to result in grandcanonical particle ratios
e.g. at midrapidity.

\noindent
How can a high multiplicity $p \overline{p}$ collision appear to be an equilibrated system?
The importance of quantum mechanical coherence for several aspects of multiparticle production
in high energy particle and nuclear 
 collisions has been discussed in \cite{dokshitzer} and  \cite{stock_review}.

\noindent
High energy particle and nuclear collisions
should not be viewed as a classical  billiard ball cascade.
Interactions among particles and virtual particle exchange
for a  quantum mechanical coherent system may
lead faster to equilibrium and therefore
to the  equilibrium particle ratios pattern which is observed for the final
(incoherent) particle species emmitted and observed in the detectors.
This is true for nuclear as well as for particle collisions at high energies. 
\\

\noindent
In the following, we address very briefly some topics of interest without going into detail.

 \subsection{ Selected topics}

\vspace{0.4cm}
 \noindent {\bf Selected topics: 1. Studies of deviations from equilibrium }

\noindent
Many interesting studies have been performed searching for deviations from 
equilibrium in particle production by introducing
new parameters like a correlation  volume  or
a factor $\gamma$ which measures deviation from equilibrium of a certain particle
type e.g. \cite{deviations,redlich,rafelski_k}.
Such studies are very interesting and promising.
\\

\noindent
However even if the introduction of free parameters like $\gamma$ factors for pions and/or strange
 particles is improving the fits,  the physics interpretation of the result may not be unique.
\\
For example the same deviation of measured ratios from a thermal
source can be interpreted as strangeness nonequilibrium, or as pion nonequilibrium
or both, or baryon nonequilibrium and so on, and have each time a better chisquare than
 a grandcanonical fit, while the deviation may even be due to something else e.g.  to
an incorrect resoncance decay correction.

\noindent
It is also unclear if the precision of the measurements and in particular
the correction for resonance decays is good enough for such studies.
The systematic errors of those corrections should be smaller than
the deviations seen in the data.
More precise data from RHIC and LHC will help such studies to really explore their potential.
\\

\noindent
On the other side, standard thermodynamics (namely when there is no distinction between
 grandcanonical ensemble and other ensembles) 
therefore with no gamma factors, allows for a conceptually 
 clear interpretation of the thermal or not thermal nature of a particle source.
It is also found that indeed  can  fit a large variety of data
 with a good chisquare \cite{pbm_first_thermal,mapping}, including
high multiplicity $p \overline{p}$ collisions \cite{mapping}. 

\noindent
 Particle yields from $p+p$ and $p \overline{p}$ collisions at highest energies are very limited. 
It is therefore very important to achieve a better understanding of hadron production and
strangeness in particular in p+p and other particle collisions
through more measurements and theoretical work like e.g. \cite{nexus}.

\vspace{0.4cm}
\noindent
{\bf Selected topics: 2. The $\phi$}

\noindent
The puzzle of different characteristics seen between the channels
 $\phi \rightarrow e^+ e^- (\mu^+ \mu^-)$
and $\phi  \rightarrow K^+ K^-$ 
possibly due to rescattering of Kaons,
is under experimental investigation  \cite{na50_phi,phenix_phi,ceres_phi,na60_phi,star_phi}.
It seems that the discrepancy in the $\phi$ yield may be reduced using a new estimated branching ratio,
however the different $p_t$ slopes still remain. No discrepancy is seen in d+Au
at RHIC by PHENIX \cite{phenix_phi}.

\vspace{0.4cm}
\noindent
{\bf Selected topics: 3. The role of Isospin corrections}

\noindent
Another interesting new experimental finding
is the fact that after correcting for the isospin the ratio $K^+/\pi^+$ in Pb+Pb over p+p
is similar to the same ratio in p+Pb over p+p \cite{rybicki_qm2004}.
It would be however interesting to see the isospin corrected overall $\lambda_s$ in full phase space or at
midrapidity, rather than only one particle ratio, to be able to draw conclusions. 
Furthermore, models could be used to study the effect of isospin correction.
Another interesting result of NA49  \cite{rybicki_qm2004} is that when loking at the enhancement of the projectile
component from the p+p expectation (p+A = p+p (0.5 Nr of collisions x alpha + 0.5 $E_{proj}$, with $E_{proj}$ the
enhancement factor and alpha the isospin correction),
  the $\Xi^-$ and $\overline{\Xi}$
have the same enhancement factor among them and  in both Pb+Pb and p+Pb collisions.
It would be important  to see the $\Xi^+$ and $\Xi^-$ ratio to participating nucleons or
to pions after isospin correction, rather than the above factor $E_{proj}$
 to be able to assess the role of this correction.

\begin{figure}
  \includegraphics[height=.26\textheight]{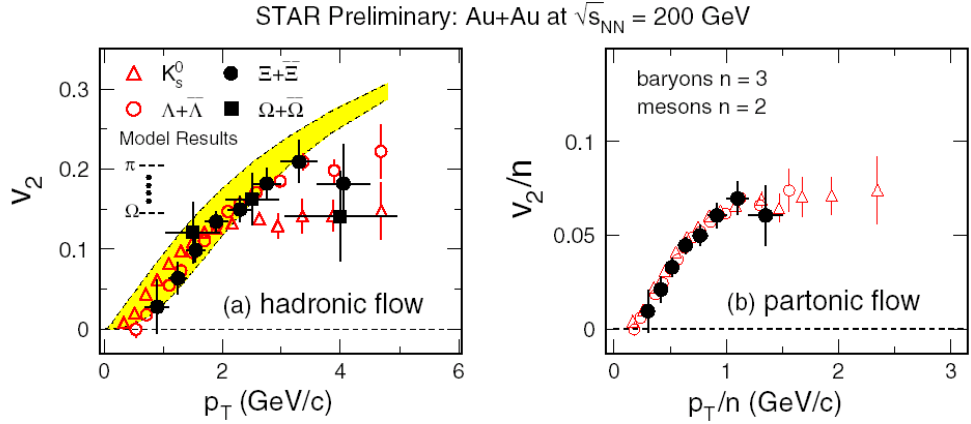}
\caption{
Left: The azimuthal anisotropy parameter v2 of strange hadrons
 as a function of $p_T$  in
minimum bias Au+Au collisions at $\sqrt{s}$=200 GeV.
The hatched band indicates results from hydrodynamical calculations.
Right:
the same while both v2 and the $p_T$ have been divided 
by the number of constituent quarks in each hadron.
}
\label{flow}
\end{figure}

\vspace{0.4cm}
\noindent
{\bf Selected topics: 4. Comparison to  coalescence}

\noindent
A further question is if yields of hadrons including strangeness
are compatible with their production by quark coalescence out of a hadronizing
QGP.
Quark coalescence models predict a universal scaling of elliptic
flow parameters versus $p_T$ with the number of constituent quarks.
There has been shown recent evidence that dividing v2 and $p_t$ by the hadron quark content
leads to a universal curve in the pt dependence of the v2 flow component 
in agreement with the above expectation (fig. \ref{flow})
 \cite{kai_qm2004}.

\section{Pentaquarks}

  \noindent
  Several models  predict the multiplet structure and characteristics of pentaquarks 
  for example the chiral soliton model, the uncorrelated quark model,  correlated
  quark models, QCD sum rules, thermal models,
  lattice QCD etc. (e.g. \cite{diakonov_polyakov_petrov_1997, navarra, jaffe, jaffe_2, octets_diakonov,ellis, guzey, stancu, glozman, aichelin_theta,stocker, 0402260}).
The current theoretical description of pentaquarks is extremely rich as well as important
and usefull for further searches, while it should be noted
that it does not lead to a unique picture
on the pentaquark existence and characteristics.
This fact reflects the complexity of the subject. 
  For example the observed mass splitting between $\Xi^{--}(1860)$ and
$\theta^+(1530)$ is unexpected or not allowed
by some authors
e.g.
 like the old predictions of the soliton model
 \cite{karlinerlipkin,diakonov_polyakov_petrov_1997} and
expected by others 
 e.g. 
 new calculations of the soliton model
\cite{ellis,weigel}.
Furthermore, lattice calculations give very different results to the questions
if pentaquarks exist  and which mass and parity they have.
The revival of the question on pentaquark existence is however very 
recent, and a fast progress is expected from both theory and experiment
in the near future.
Here we will concentrate on the experimental results.

\vspace{0.4cm}
\noindent
{\bf $\theta^+_{ \overline{s} }  $}

\noindent
 The first prediction
  of the mass of the state $uudd \overline{s}$ \cite{pra}
   using the chiral soliton model 
  was 
   m($uudd \overline{s}$) = 1530 MeV.
  A recent updated study of pentaquarks within this model
  has shown that this value has a systematic error of the order of $\pm$ 100 MeV
  \cite{ellis}.

\begin{figure}
  \includegraphics[height=.32\textheight]{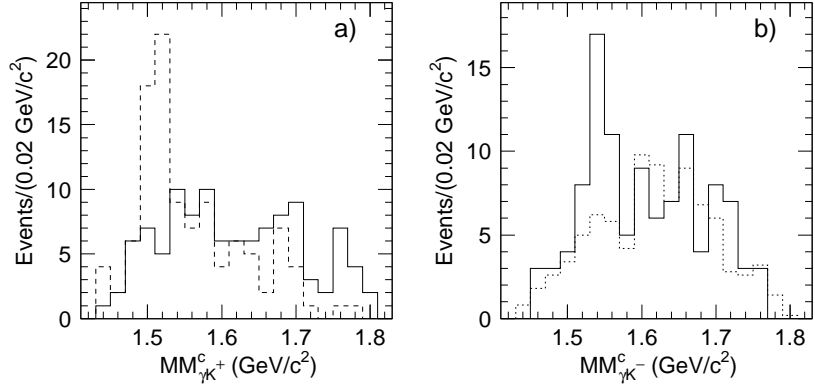}
  \caption{
Missing mass $M(\gamma,K^+)$  and right of the  $M(\gamma,K^-)$ (right)
measured by the LEPS Collaboration. See text for explanations.
}
\label{nakano}
\end{figure}

 \noindent
  Recent
  advances in theoretical description of pentaquark characteristics,
  in particular the prediction of the width of $\theta^+$( $uudd \overline{s}$) 
  with spin=1/2 to be below 15 MeV 
  \cite{diakonov_polyakov_petrov_1997}
 as well as in experimental methods and instrumentation \cite{nakano}
  lead to
 the observation of the  $\theta^+$( $uudd \overline{s}$)
 in the $\gamma n \rightarrow  \theta^+ K^-   \rightarrow K^+ n K^- $
 reaction
 by the LEPS collaboration.
 They used $\gamma$ beam with energy 1.5-2.4 GeV on
 C and H targets to be able to study  $\gamma n$ and $\gamma  p$ reactions.
 They found m( $\theta^+$) = 1540 $\pm$ 10 $\pm$ 5 (syst) MeV,
 width less than 25 MeV and $S/ \sqrt{B}$ = $19/\sqrt{17}$= 4.6.
 The
  neutron was inferred by missing mass measurement.
 Many systematic studies have
 been performed e.g. for the understanding of the background.
 Figure \ref{nakano}
 shows left the missing mass $M(\gamma,K^+)$, and right of the  $M(\gamma,K^-)$. 
 Left the dashed line shows the $\Lambda (1520)$ peak when a proton has been detected
 due to 
  $\gamma p \rightarrow   K^+ K^- \Lambda (1520) $.
 The solid line shows the $\theta^+$ selected sample after all cuts, which does not exhibit a 
 $\Lambda(1520)$ peak. Therefore, the "signal" sample is dominated by $\gamma n $ interactions.
 The right figure shows the missing mass $M(\gamma,K^-)$ 
 for the  $\theta^+$ selected sample after all cuts
 again as solid line and the background with dashed.
 The $\theta^+$ peak is visible.
 Among the systematic studies performed were to intentionally missidentify pions as kaons,
 test if the tails of the $\phi \rightarrow K^+ K^-$ distribution generate
 a peak, test  if stronger particle identification destroyes the peak, Monte Carlo studies,
 test if $\Lambda$ and $\Sigma$ peaks are well reproduced etc.
\\
Most importantly,
recent preliminary analysis of figures.data taken recently by LEPS lead to a confirmation
of the seen peak with about 90 entries in the peak above background, as compared
to 19 measured previously \cite{leps_pentaquark2004}.

\noindent
This first observation were followed by a number of experiments which have seen
the $\theta^+$ peak.
The DIANA collaboration 
at ITEP (bubble chamber experiment)
  used $K^+$ beam with energy 850 MeV on
Xe and
have observed a peak in the invariant mass $K^0_s p$ from the
 reaction
$ K^+ Xe \rightarrow K^0_s p Xe^{'} $ 
\cite{diana}.
 They found m( $\theta^+$) = 1539 $\pm$ 2 MeV,
 width less than 9 MeV and $S/ \sqrt{B}$ = 4.4.

\begin{figure}
  \includegraphics[height=.3\textheight]{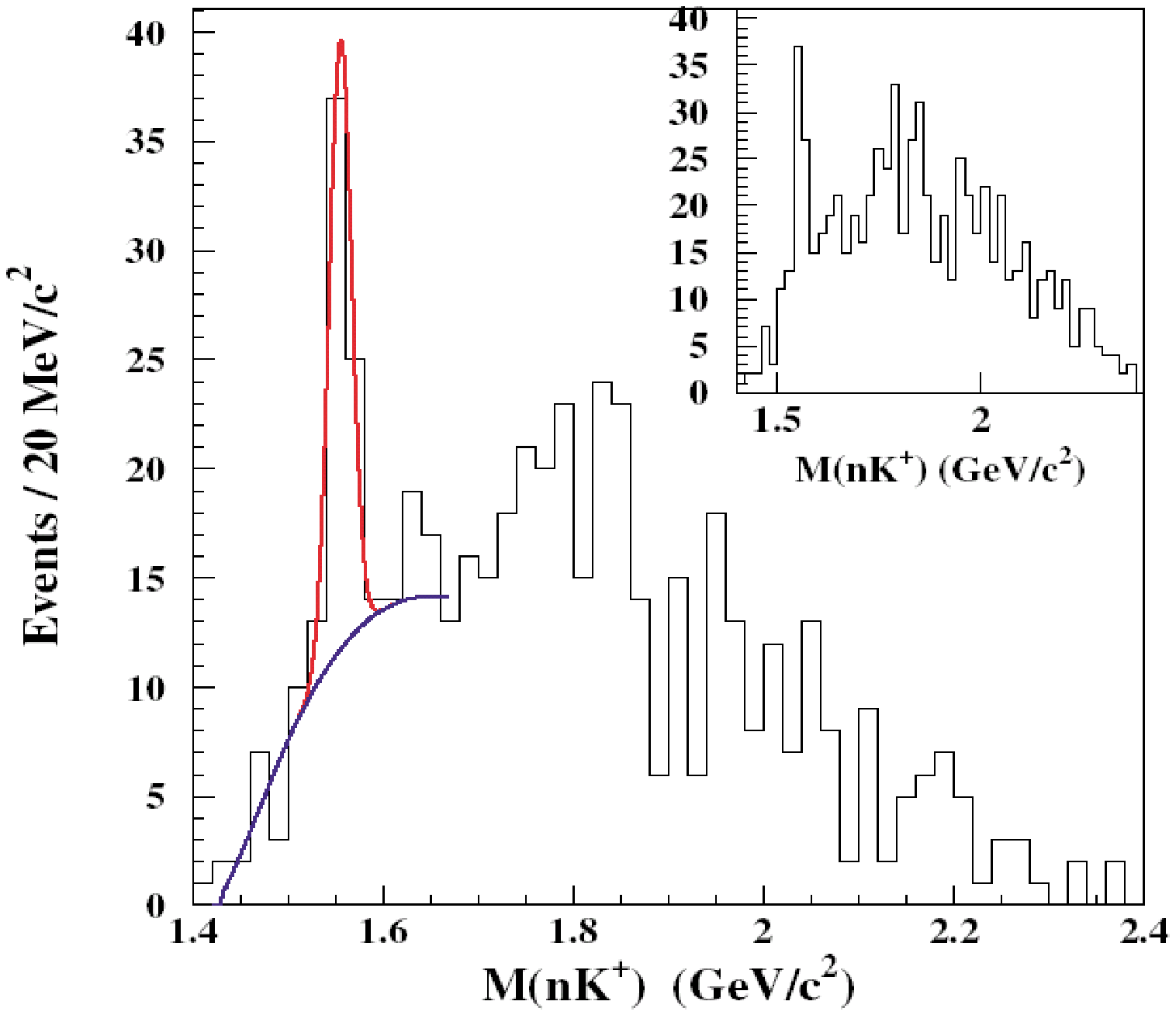}
\hspace*{0.5cm}
\caption{Invariant mass $n K^+$ 
measured by the CLAS Collaboration. See text for explanations.}
\vspace*{-0.4cm}
\label{clas_nstar}
\end{figure}

\begin{figure}
  \includegraphics[height=.33\textheight]{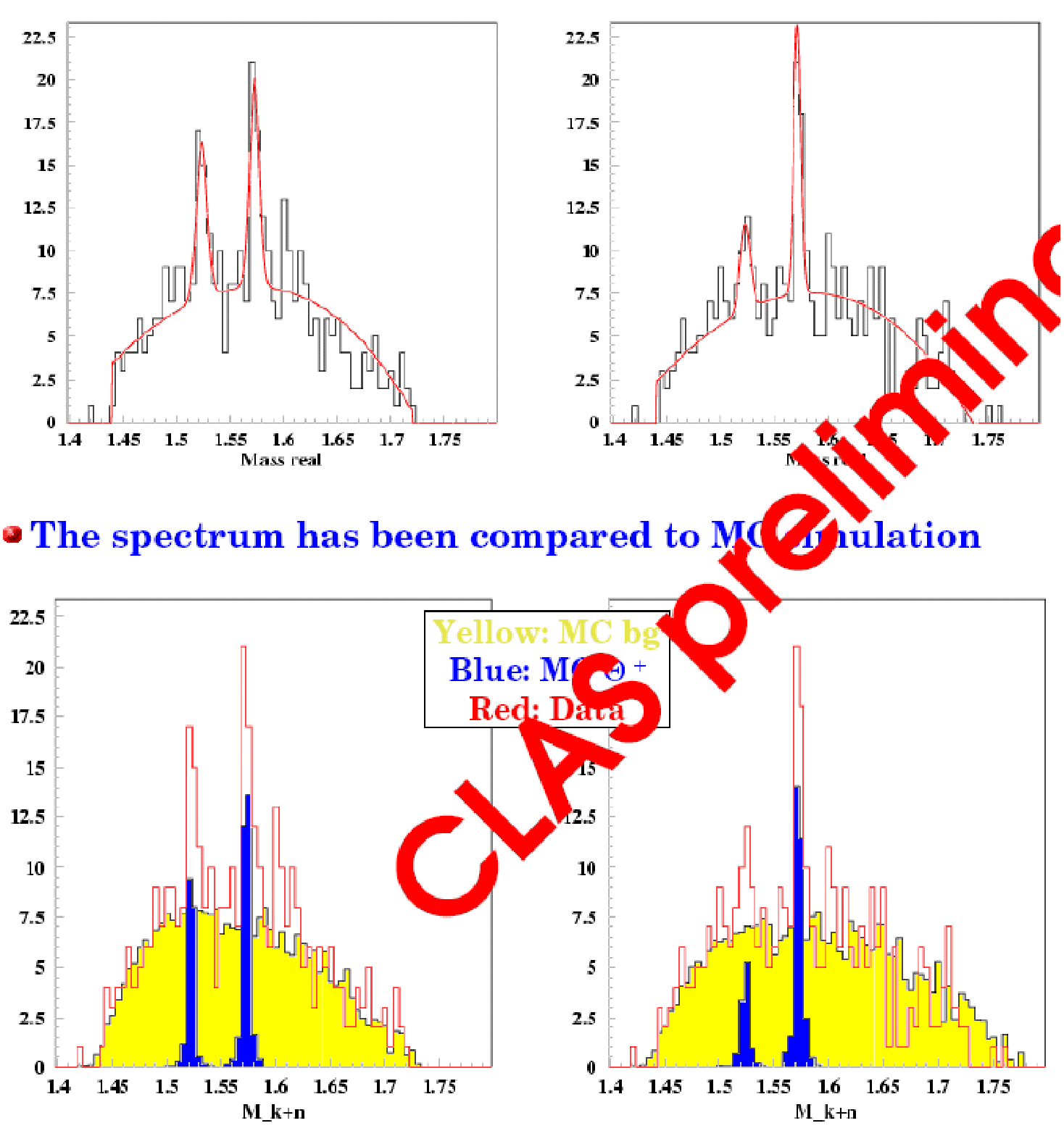}
\caption{Invariant mass $n K^+$ 
measured by the CLAS Collaboration. See text for explanations.}
\vspace*{-0.4cm}
\label{clas_bata}
\vspace*{-0.4cm}
\end{figure}

\noindent
The SAPHIR  collaboration 
at ELSA
  used $\gamma$ beam with energies 31-94\% of 2.8 GeV on
H and
have observed a peak in the invariant mass $n K^+$ from the
 reaction
$ \gamma p \rightarrow \theta+ K^0_s  \rightarrow n K^+ K^0_s $ 
\cite{saphir}.
 They found m( $\theta^+$) = 1540 $\pm$ 4 $\pm$ 2 MeV,
 width less than 25 MeV and $S/ \sqrt{B}$ = 5.2.

\noindent
The HERMES  collaboration 
at DESY
  used $e^+$ beam with energy 27.6 GeV on
deuterium and
have observed a peak in the invariant mass $p K^0_s$
\cite{hermes}.
 They found m( $\theta^+$) = 1528 $\pm$ 2.6 $\pm$ 2.1 MeV,
 width 17 $\pm$ 9 $\pm$ 3 MeV and $S/ \sqrt{B}$ = 4.2 to 6.3.
While the signal to background ratio in the above publication is 1:3,
new analysis lead to an  improved signal to background ratio of 2:1
\cite{hicks_pentaquarks2004}.

\noindent
The COSY-TOF  collaboration 
 observed a peak in the invariant mass $p K^0_s$
 from the reaction $ p  p 
	  \rightarrow \Sigma^+ \theta^+
	  \rightarrow ( n \pi^+) ( K^0_s p)
    $
\cite{cosytof}.
 They found m( $\theta^+$) = 1530 $\pm$ 5 MeV,
 width below 18 $\pm$ 4 MeV and $S/ \sqrt{B}$ = 5.9.
They measure a cross section of 0.4 $\pm$ 0.1 $\pm$ 0.1 (syst) $\mu b$
which is in rough agreement with predictions of
0.1-1 $\mu b$
for p+p, p+n near threshold.

\noindent
An analysis of old $\nu$ and $\overline{\nu}$ interactions
from old bubble chamber experiments
filled with H, d, or neon,
and beam energies of 40 or 110 GeV
has resulted in a $\theta^+$ peak in the 
invariant mass $p K^0_s$
with
    m( $\theta^+$ ) = 1533 $\pm$ 5 MeV,
 width less than 20  MeV
and $S/ \sqrt{B}$= 6.7
\cite{neutrino}.

\begin{figure}
  \includegraphics[height=.35\textheight]{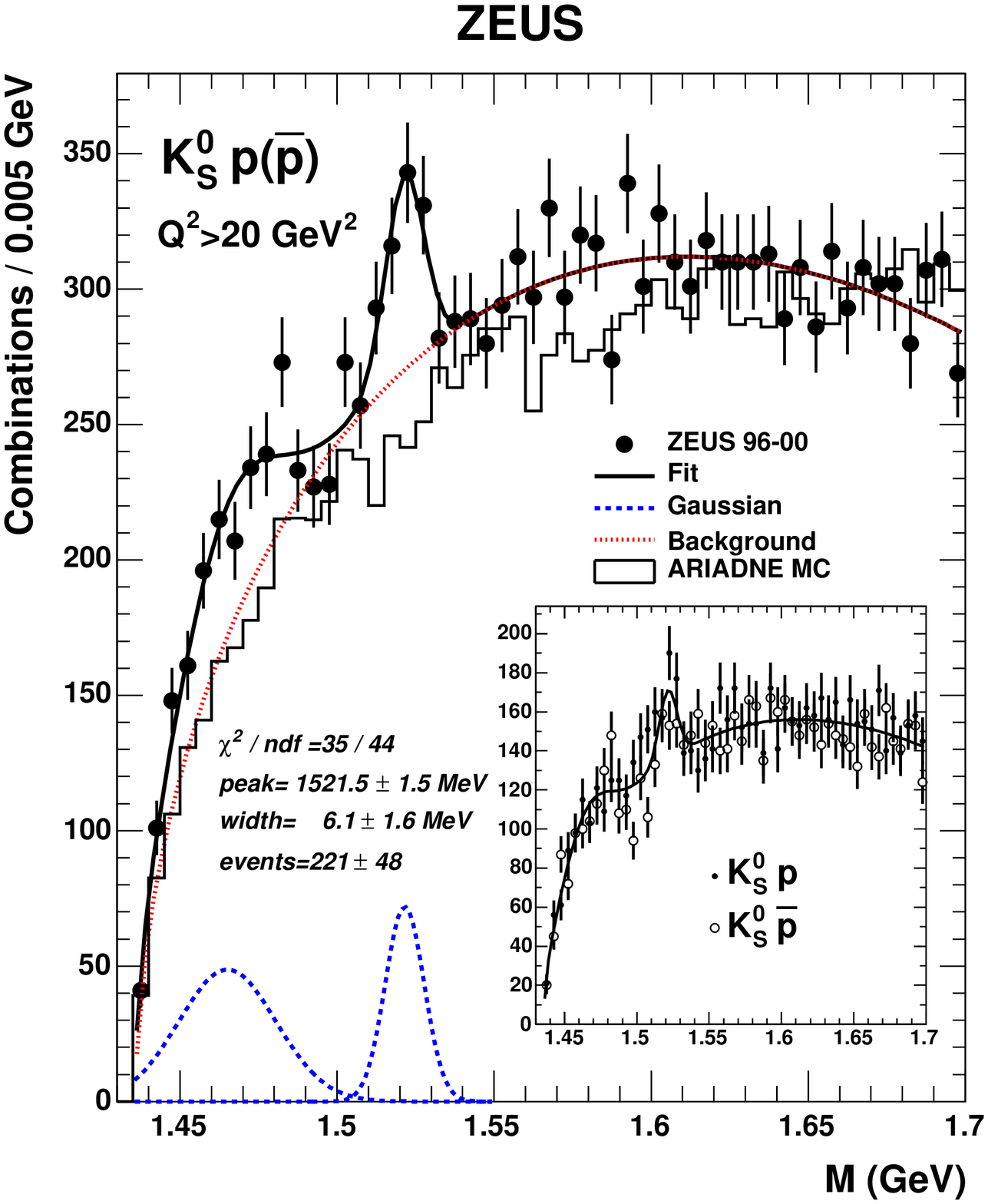}
  \includegraphics[height=.45\textheight]{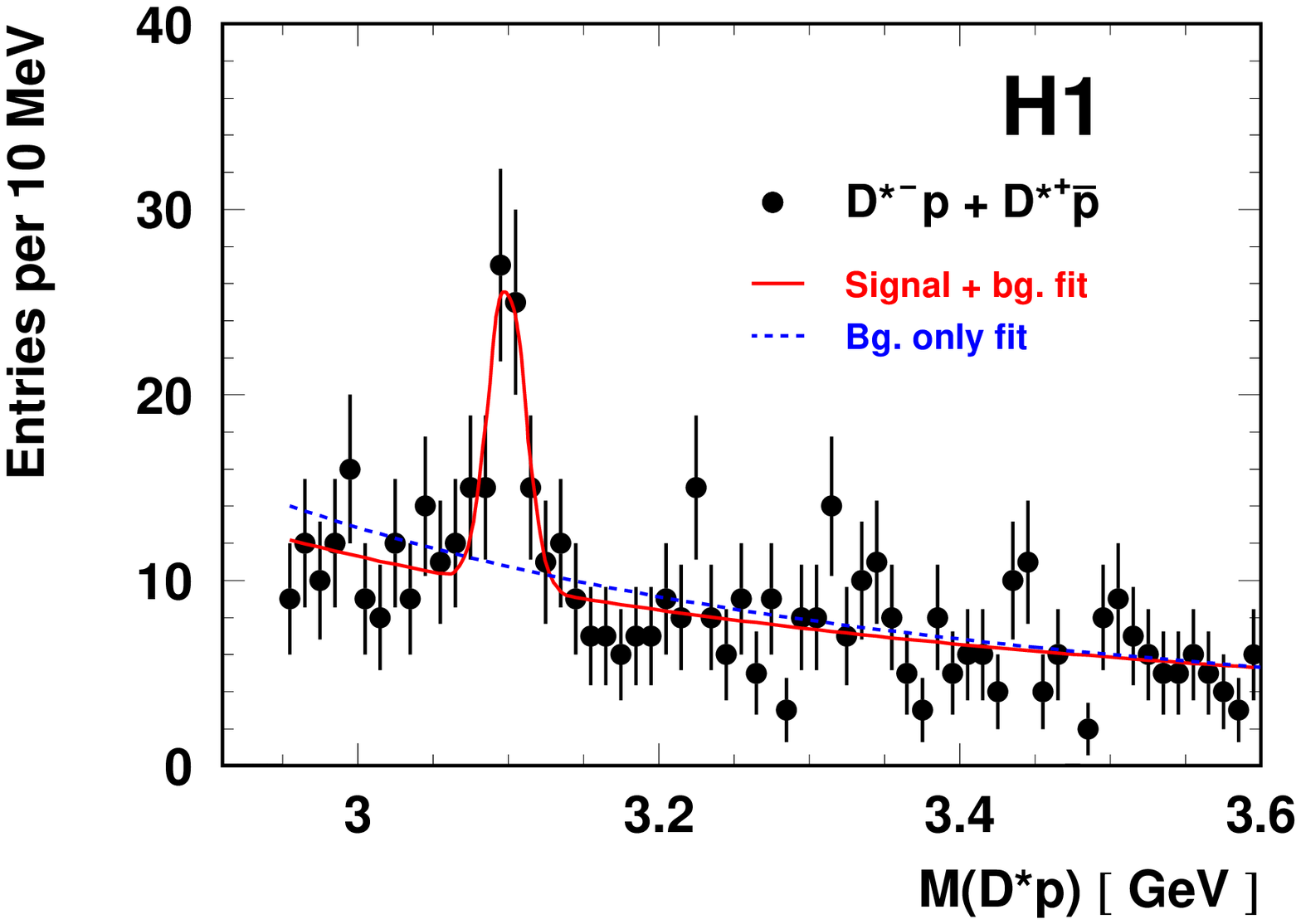}
\caption{Left: Invariant mass $K^0_s p(\overline{p})$ measured by the Zeus Collaboration. 
Right: Invariant mass $ D^{*-} p$ and $ D^{*+} \overline{p}$
measured by the H1 collaboration.}
\label{zeus_hermes}
\end{figure}

\noindent
The CLAS  collaboration 
  used $\gamma$ beam with energy about 95\%(2.474-3.115) GeV 
on deuteron target
and 
have observed a peak in the invariant mass $n K^+$
from the reaction 
$ \gamma d \rightarrow K^+ K^- p n 
$
through missing mass measurement of the neutron
\cite{clas_1}.
 They found    m( $\theta^+$) = 1542 $\pm$ 5 MeV,
 width of 21 MeV consistent with the experimental resolution,
  and $S/ \sqrt{B}$ = 5.2 $\pm$ 0.6.

\noindent
In a later publication the CLAS  collaboration 
  used $\gamma$ beam with energy 3-5.47 GeV 
on deuteron target
and 
have observed a peak in the invariant mass $n K^+$
from the reaction 
$ \gamma p \rightarrow \pi^+ K^- K^+ n 
$
through missing mass measurement of the neutron
(fig. \ref{clas_nstar})
\cite{clas_2}.
They observed evidence that the $\theta^+$ is preferably
produced through the decay of a new narrow resonance $N^0(2400)$.
 They found    m( $\theta^+$) = 1555 $\pm$ 10 MeV,
 width less than 26 MeV
  and $S/ \sqrt{B}$ = 7.8 $\pm$ 1.
This is the highest published significance obtained for the $\theta^+$ from
a single measurement.

\noindent
A preliminary analysis of CLAS
of the reaction
$\gamma d 
	\rightarrow \theta^+ \overline{ K^0} 
	\rightarrow (K^+ n) K^0_s$
 \cite{trento_bata}
(fig. \ref{clas_bata})
  shows
two peaks in the invariant mass ($K^+ n$)
 at 1523 $\pm$ 5 MeV and at 1573 $\pm$ 5 MeV
both having a width of about 9 MeV
and significance of 4, respectively of 6 $\sigma$.
It is important that CLAS will clarify the reason for the shift of the
lower mass peak position and why the second peak appears with the new cuts
but not with the old ones in the previously studied reactions.
The second peak is a candidate for an excited $\theta^+$ state which is expected 
to exist with about $\sim$ 50 MeV higher mass than the ground state, in
agreement with the observation.
A preliminary cross section estimate gives 5-12 nb for the low mass peak
and 8-18  nb for the high mass peak.
\noindent
The above two peaks have been quoted also in \cite{bata_pentaquarks2004}.

\noindent
CLAS has taken a large amount of data in 2004 which are now been analysed.
First results have been quoted 
which confirm the previous $\theta^+$ observations with new peaks in different
channels, all near 1.55 GeV
 \cite{clas_pentaquarks2004}.

\noindent
The ZEUS  collaboration 
at DESY
  used $e^+p$ collisions at $\sqrt{s}$=300-318 GeV 
 and
have observed a peak in the invariant mass $p K^0_s$
(fig. \ref{zeus_hermes}, left)
  \cite{zeus}.
They have observed for the first time the $\overline{ \theta}^-$
state decaying in $\overline{p} K^0_s$.
 They found    m( $\theta^+$ + $ \overline{ \theta}^-$) = 1527 $\pm$ 2 MeV,
 width 10 $\pm$ 2 MeV.

\noindent
The NA49 experiment has also reported recently
a preliminary result of a peak observed in the invariant mass $p K^0_s$ 
in p+p reactions at $\sqrt{s}$=17 GeV
with mass 1526 $\pm$ 2 MeV and width below 15 MeV
\cite{kadija_pentaquark2004}.
They have also reported a preliminary evidence that the
$\theta^+$ peak appears more pronouned after
assuming $\theta^+$ production through the decay of a resonance
$N^0(2400) \rightarrow \theta^+ K^-$ \cite{barna}
as suggested by CLAS \cite{clas_2} and discussed in
\cite{karliner,azimov_n2400}.

\noindent
NOMAD has shown recently 
preliminary results on the observation of a 
$\theta^+$ candidate peak in the $p K^0_s$ invariant mass
using their full statistics of $\nu A$ interactions, with a mean energy of 
the $\nu$ beam of 24.3 GeV \cite{camilleri}.
The mass observed is 1528.7 $\pm$ 2.5 MeV
and the width is consistent with the 
experimental resolution of 9 MeV.

\noindent
GRAAL has shown preliminary results on the observation of a 
$\theta^+$ candidate peak in the $p K^0_s$ invariant mass
in 
 $\gamma d \rightarrow \theta^+ \Lambda^0 \rightarrow (K^0_s p) \Lambda^0$
 interactions, using $\gamma$ energy of maximally 1.5 GeV \cite{graal_pentaquark2004}.
The mass observed is 1531 MeV while no error is given.
\\

\noindent
Most of the experiments measure a $\theta^+$ width consistent with the experimental resolution,
while few of them give a measurement of width somewhat larger than their resolution
namely Zeus and Hermes.
A measurement with a much improved resolution would be important.
\\
Non-observation of $\theta^+$ in previous experiments lead to an estimate
of its width to be of the order of 1 MeV or less \cite{1mev_arndt}.
This limit would gain in significance, once the $\theta^+$
non-observation by several experiments will be better understood,
excluding other reasons for the $\theta^+$ non-observation in the examined
reactions.

\vspace*{0.4cm}
\noindent
{\bf Do the measured $\theta^+$ masses vary as expected for a real state ?}

\begin{figure}
  \includegraphics[height=.42\textheight]{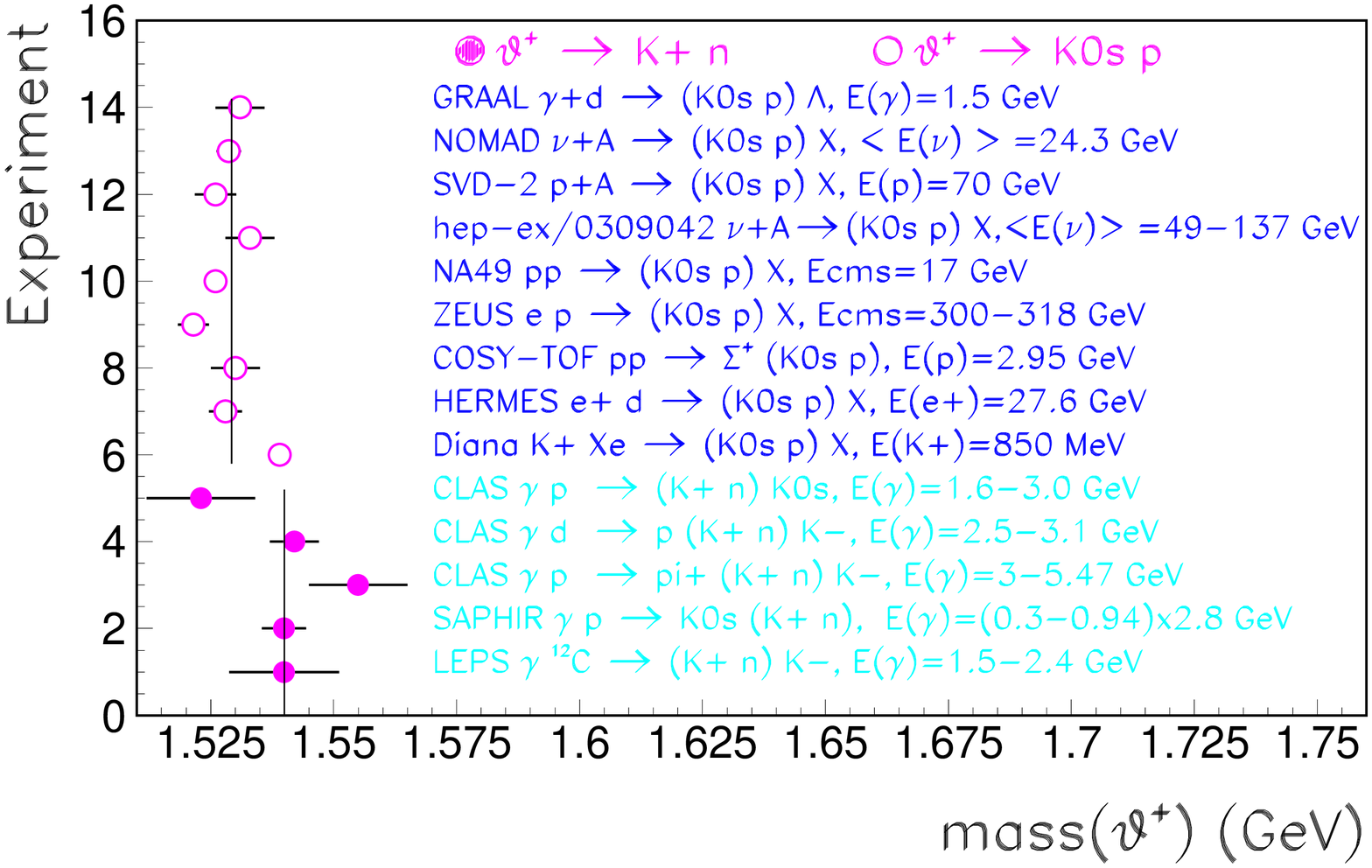}
\vspace*{-0.4cm}
\vspace*{-0.4cm}
\vspace*{-0.3cm}
  \caption{
Compilation of measured $\theta^+$ masses.
}
\label{theta_mass}
\vspace*{-0.3cm}
\end{figure}

\noindent
Figure \ref{theta_mass} shows a compilation of the 
masses of $\theta^+$ candidate peaks observed
by several experiments.
The statistical and systematic errors (when given) have been added in quadrature.
For GRAAL 
we assume an error of 5 MeV as no error has been given in \cite{graal_pentaquark2004}.
For the two preliminary peaks of CLAS we assume the systematic error
of 10 MeV quoted previously by CLAS.
The lines indicate the mean value of the mass among 
the $\theta^+ \rightarrow p K^0_s$
and the
$\theta^+ \rightarrow n K^+$
observations.
It appears that the 
mass of $\theta^+$
from $\theta^+ \rightarrow n K^+$
observations
is systematically higher than the one
from 
$\theta^+ \rightarrow p K^0_s$
observations.
This may be related to the special corrections needed
for the Fermi motion and/or to details of the
analysis with missing mass instead of direct measurement 
of the decay products.

\noindent
All observations together give a mean mass of 
1.533 $\pm$ 0.023 GeV and they
deviate from their mean with a
 $\chi^2/DOF$
of 3.92.
The $\chi^2/DOF$ for 
the deviation of the $\theta^+ \rightarrow p K^0_s$
observations
from their mean of 1.529 $\pm$ 0.011 GeV
is 
3.76.
The $\chi^2/DOF$ for 
the deviation of the $\theta^+ \rightarrow n K^+$
observations
from their mean of 1.540 $\pm$ 0.020 GeV
is 
is 0.94.

\noindent
The bad $\chi^2/DOF$ for the $\theta^+ \rightarrow p K^0_s$
observations
maybe due to an underestimation of the systematic errors.
In particular in some cases no systematic errors are given,
sometimes because the results are preliminary.
If we add a systematic error of 0.5\% of the measured mass (therefore
of about 8 MeV) on all
measurements for which  no systematic error 
was given by the experiments,
we arrive to a
$\chi^2/DOF$ for the $\theta^+ \rightarrow p K^0_s$
observations
of 0.95
and a mean mass of 1.529 $\pm$ 0.022 GeV.
The $\chi^2/DOF$ for the $\theta^+ \rightarrow n K^+$
observations
almost don't change by this,
(mean mass = 1.540 $\pm$ 0.022 GeV, $\chi^2/DOF$=0.91),
because the experiments mostly give the systematic errors
for this decay channel.
All observations together give then a mean mass of 
1.533 $\pm$ 0.031 GeV and they
deviate from their mean with a
 $\chi^2/DOF$
of 2.1, reflecting mainly the difference of masses between
the two considered decay channels.
It is important to understand the origin of this discrepancy.
\\
This problem can be studied measuring $\theta^+ \rightarrow K^+ n$ in experiments
with direct detection of the neutron or the antineutron for the $\overline{ \theta^-}$
 like PHENIX and GRAAL.

\vspace*{0.4cm}
\noindent
{\bf $\theta^{++}$ }

\noindent
A preliminary peak is quoted by CLAS \cite{bata_pentaquarks2004}
 for the candidate $\theta^{++} \rightarrow
p K^+$ produced in the reaction 
$ \gamma p \rightarrow \theta^{++} K^-
		\rightarrow p K^+ K^-$
 at 1579 $\pm$ 5 MeV.
A previous peak observed by CLAS in the invariant mass $ p K^+$
has been dismissed as due to $\phi$ and hyperon resonance reflexion
\cite{clas_thetapp_1}.
\\
\noindent
The STAR collaboration quoted a preliminary peak in the $p K^+$ and 
$\overline{p} K^-$ invariant
masses at 1.530 GeV, with $S/\sqrt{B}$ $\sim$ 3.8 and width about 9 MeV,
  which is candidate
for the
 $\theta^{++} \rightarrow p K^+$
as well as the antiparticle
 $\overline{ \theta^{--}}  \rightarrow \overline{p} K^-$
in d+Au collisions at $\sqrt{s}$=200 GeV \cite{star_thetapp}.
\\
HERMES has reported the non-observation of a peak 
in the $p K^+$ invariant masses \cite{hermes}.

\vspace*{0.4cm} 
\noindent
{\bf $\Xi$, $N^0$}

\begin{figure}
  \includegraphics[height=.35\textheight]{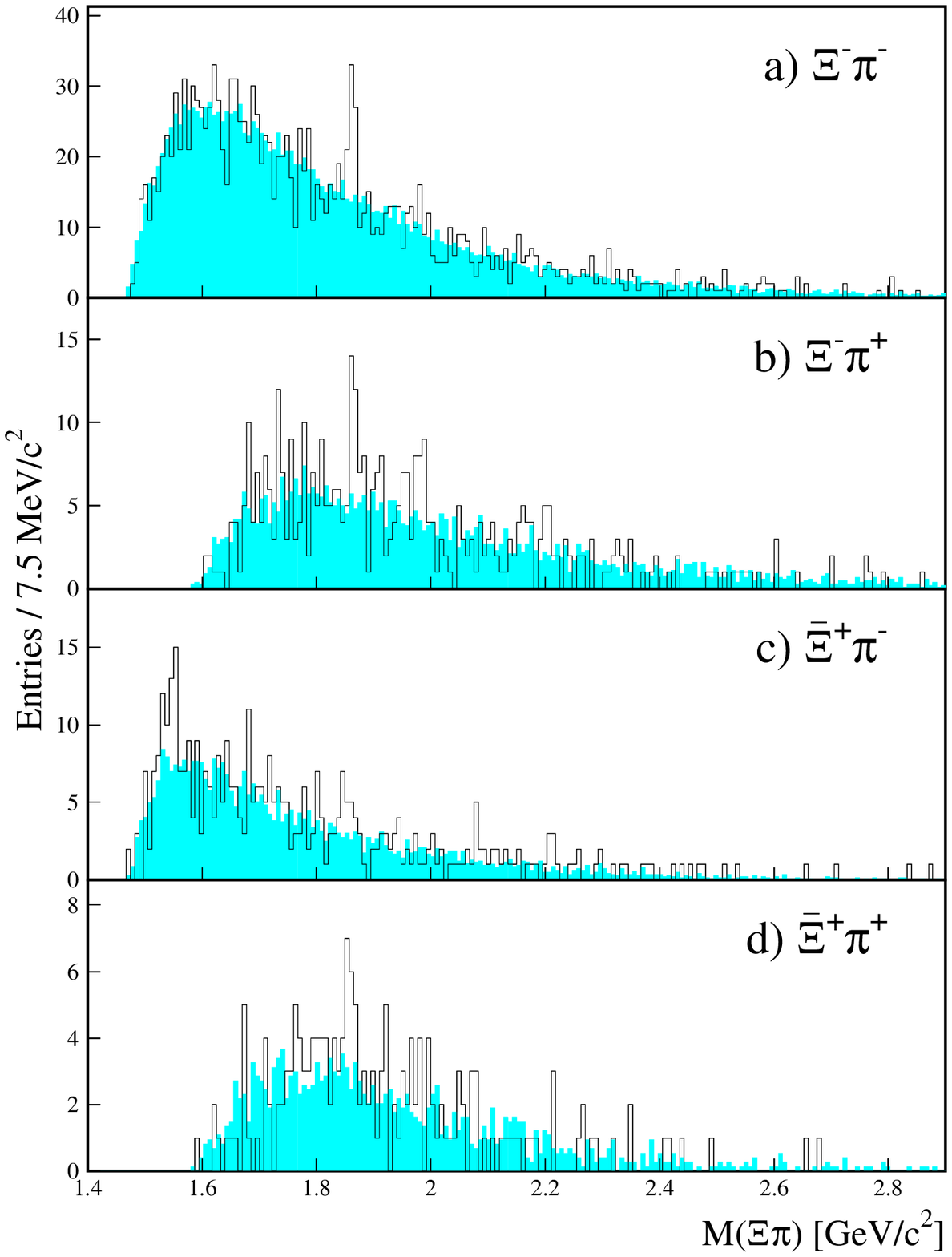}
  \includegraphics[height=.35\textheight]{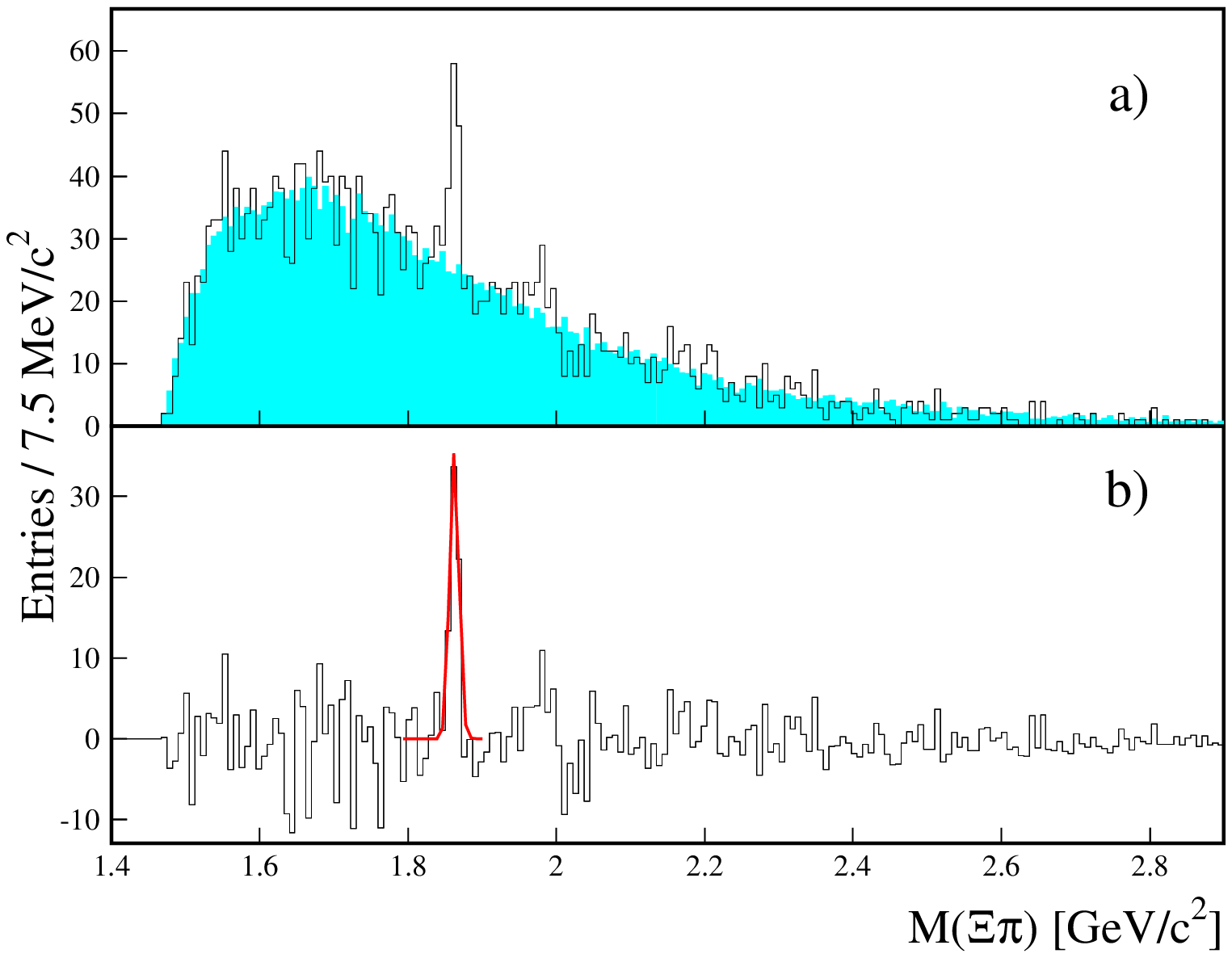}
  \caption{Invariant mass $\Xi^- \pi^-$, $\Xi^- \pi^+$ and their anti-channels
 and their total sum measured
by the NA49 collaboration.}
\label{ximinmin}
\end{figure}

\noindent
The NA49 experiment has observed in p+p reactions
at $\sqrt{s}$=17 GeV 
the pentaquark candidates 
$\Xi^{--}(1862 \pm 2  MeV) \rightarrow \Xi^- \pi^-$,
the
$\Xi^{0}(1864 \pm 5  MeV) \rightarrow \Xi^- \pi^+$
and their antiparticles  \cite{na49}.
They measure a width consistent with their resolution of about  18 MeV.
They also observe preliminary results of the decay
$\Xi^-(1850) \rightarrow \Xi^0(1530) \pi^-$
(fig. \ref{ximinmin})
with simarly narrow width as the other candidates \cite{na49_ximinus}.
The $\Xi^{--}$ is a candidate for the antidecuplet 
and the $\Xi^0$ too due to the small mass difference
while it is unclear if the $\Xi^-$(1850) is 
from the octet or the antidecuplet.
An observation of the $\Xi$ I=1/2 from the octet in $\Lambda K^0_s$
would answer this question.
 The non observation of a peak in the invariant mass $\Xi^0(1530) \pi^+$
 is also an important information, as this decay channel
 is not allowed by SU(3)
 for the antidecuplet $\Xi^+$ \cite{jaffe_2}.

\begin{figure}
  \includegraphics[height=.32\textheight]{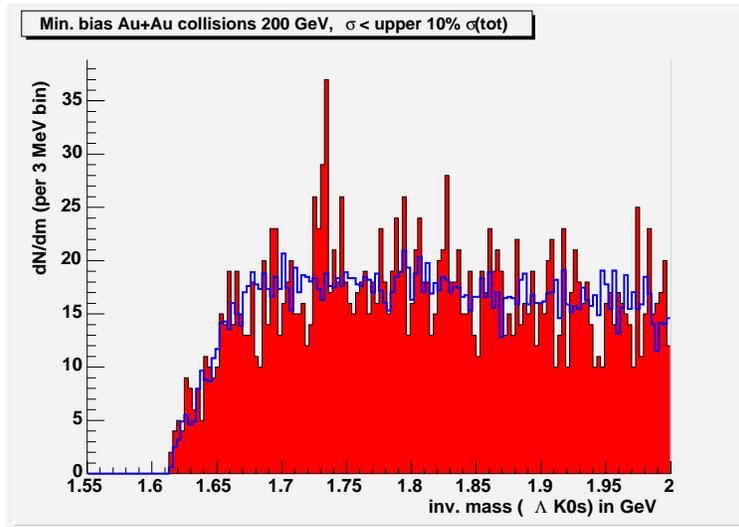}
  \caption{Invariant mass $\Lambda K^0_s$ measured by the STAR experiment. See text for explanations.}
\label{n0}
\end{figure}

\noindent
The experiment STAR has shown preliminary results on a 
$N^0$ ($udsd \overline{s}$, $udd u \overline{u}$)
 or $\Xi$ ($uds s \overline{d}$) I=1/2 candidate.
The $N^0$ can be a mixture of the quark contents
($udsd \overline{s}$, $uddu \overline{u}$).
STAR uses minimum bias Au+Au collisions at $\sqrt{s}$=200 geV
and  observes a peak in the decay channel
	$\Lambda K^0_s$ at a mass 1734 $\pm$ 0.5 (stat) $\pm$ 5 (syst) MeV
(fig. \ref{n0})
with width consistent with the experimental resolution of about  6 MeV
and $S/ \sqrt{B}$ between 3 and 6 depending on the method used
\cite{jamaica}.
Extensive systematic studies have been performed, investigating
particle
missidentifications, split tracks and kinematic reflexions. 
They also observe signs of the known narrow states
$\Xi(1690)$ and $\Xi(1820)$ with a lower significance.
They don't observe a peak near 1850 - 1860 MeV 
 resulting from a $\Xi^0$ I=1/2 (octet) pentaquark state
with the same mass as the $\Xi^-(1850)$ of NA49,
disfavouring the picture of degenerate octet and antidecuplet 
even though a low branching ratio to $\Lambda K^0_s$
may not allow to observe the peak with the present statistics.

\noindent
The GRAAL experiment has shown preliminary results on 
two narrow  $N^0$ candidates.
One candidate is observed at a mass of 1670 MeV in  the invariant mass of
$ \eta n$ from the reaction
$ \gamma d \rightarrow \eta n X$.
 The neutron has been directly detected.
The other is observed at a mass of 1727 MeV
 in the invariant masses of 
$\Lambda K^0_s$  as well as 
in the invariant masses of
$\Sigma^- K^+$
at the same mass and with the same width
\cite{trento_graal}.
The second reaction allow to establish the strange quark content  and
therefore to exclude the $\Xi$ hypothesis.
The difference of 7 MeV between the STAR and GRAAL 
measured masses of 1727 and 1734 MeV, should be compared to the
systematic errors. STAR quotes a systematic error of 5 MeV
while GRAAL quotes no systematic error.

\noindent
The mass of the peaks  at 1670 and at  (1727,1734) MeV  is in  good agreement with the
$N$ masses 
  suggested by Arndt et al \cite{arndt0312126}. In this paper
a modified Partial Wave analysis allows to search for narrow
states and presents two candidate $N$ masses, 1680 and/or 1730 MeV
with width below 30 MeV.

\noindent
While the above mentioned narrow $N^0$ candidates 
of mass 1670 and 1727-1734 MeV
fit well into the picture of the expected $N$ and $N_s$ pentaquark
candidates, they can also be something else than pentaquarks,
 e.g. a new $N^0$=$udd$ resonance, or a $udd gg$ state.
This statement is  true  for all non-exotic pentaquark candidates.

\vspace{0.4cm} \noindent {\bf $\theta^0_{\overline{c}}$}

\noindent
The H1  collaboration 
at DESY
  used $e^- p$ collisions at $\sqrt{s}$=300 and 320 GeV 
 and
have observed a peak in the invariant masses
$ D^{*-} p$
and
$ D^{*+} \overline{p}$
(fig. \ref{zeus_hermes}, right)
at a mass 3099 $\pm$ 3 (stat) $\pm$ 5 (syst) MeV
and width of 12 $\pm$ 3 MeV  \cite{h1}.
The probability that the peak is a background fluctuation is less than $4 \ 10^{-8}$.
Extensive systematic studies have been performed.
The momentum distribution of the signal is as expected for a particle at this mass.
This peak is a candidate for the state $\theta^0_{\overline{c}} $ = $uudd \overline{c}$
and is the first charmed pentaquark candidate seen.
The mass and width of the particle and it's antiparticle are consistent.

\vspace{0.4cm} \noindent
{\bf Non-observations}

\noindent
Several experiments have reported preliminary results on the non-observation
of pentaquarks e.g. $e^+e^-$: Babar, Belle, Bes, LEP experiments, 
$p \overline{p}$: CDF, D0 $pA$:E690, $eA$: HERA-B, $ep$: Zeus (for the
$\theta^0_c$) \cite{non}.
\\
It has been  argued that the non-observation of pentaquark
states in the above
experiments is due to an additional strong suppression factor
for pentaquark production
in $e+ e^-$ collisions, as well as in B decays 
which is lifted in reactions like $\gamma A$
 in which $s \overline{s}$ and a baryon  are present in the initial state
\cite{karliner}.
The constituents of the $\theta^+$ are already present in the initial
state of e.g.  low energy photoproduction experiments, while
in other experiments baryon number and strangeness must be created from gluons \cite{karliner}.
The penalty for their production could be estimated from data e.g.
$\overline{d} / \overline{p}$ ratios \cite{karliner}.
It is important to try to assess the expected cross sections \cite{wu}.
\\
The CDF ($p \overline{p}$) and E690 (pA) non observation of pentaquarks
can be a consequence of the decrease of the pentaquark cross section
with increasing energy \cite{0406043,titov}.
This depends however on the kinematic region considered, and 
it is suggested to look for pentaquarks in the central rapidity region  \cite{0406043,titov}.
\\
In addition, if the $\theta^+$
is produced preferably through the decay of a new resonance 
$N^0(2400) \rightarrow \theta^+ K^-$ 
as suggested by CLAS and NA49 and as discussed in \cite{karliner,azimov_n2400},
neglecting this aspect maybe a further cause of its non-observation in some experiments.
\\
Some authors point out the importance to 
exclude kinematic reflexions as reason behind the $\theta^+$ peak
\cite{dzierba}.
This known source of systematic errors is investigated by the experiments
which observe pentaquark candidates.
Other authors discuss limits from the non observation
of the $\Xi(1860)$ peaks of NA49 by previous experiments \cite{wenig}.
These points are addressed while trying to explain the
new non-observations of pentaquarks quoted above.
\\

\noindent
There were also rumors spread around in the physics community 
in particular 
e.g.  a  $e^+e^-$ experiment  has found that a certain error in the identification
of particles can  lead to a peak in the $D^{*-} p$  invariant mass
near 3.1 GeV faking the $\theta^0_c$ candidate of H1.
However H1 is  aware of this fact and at which conditions it appears, and the
fake peaks have been carefully excluded.
This shows that it is important to document all findings and examine all evidence.
\\
It can certainly happen that a systematic error can lead to a peak
coinciding with the real one or the expected one, in which case it may be missinterpreted.
Another example is PHENIX which
 presented a peak at 1.540 GeV in the $\overline{n} K^-$ invariant mass,
 candidate for the $\overline{\theta^-}
\rightarrow \overline{n} K^-$ which has been later understood to be due to
a calibration error \cite{phenix}.

\vspace*{0.2cm}
\noindent
It is clear that a higher statistics is desirable in order to confirm
the pentaquark observations reported so far, as well as more measurements
and searches by other experiments.
New data taken in 2004 and planned to be taken in 2005 will
lead to enhancements in statistics of experiments up to a factor of 10
allowing to test the statistical significance and make more systematic studies.
Experiments searching for pentaquarks should  test also the production mechanisms 
proposed in the literature e.g. the $\theta^+$ production
through the $N^0(2400)$ decay.
For example
 Phenix could  search for the final state
$ \overline{ \theta^-} K^+ $
or 
$ \overline{ \theta^-} K^0_s $
demanding the invariant mass of $ \overline{ \theta^-} K^0_s $ and
$ \overline{ \theta^-} K^+ $
to be in the range 2.3 to 2.5 GeV,
and study the option to trigger online on this channel.
\\
Furthermore, not
 only the observations of pentaquark candidates but also
the  non-observations  should be published
and be well documented
in order to facilitate a better understanding of the underline physics.

\section{Summary and conclusions}

\subsection{Strangeness}

\noindent
An enhancement in the
 strangeness to pion ratio  has been observed in A+A collisions with respect to p+p and p+A
collisions at the same energy.
A spectacular enhancement has been measured especially in strange baryon and antibaryon
production per participant nucleon (N) in A+A collisions at SPS and RHIC as compared
to p+A collisions at the same energy.
However
this enhancement increases with decreasing energy, disqualifying it as QGP signature.
This behaviour can  be understood as due to the fact that at the same energy p+p , p+A and A+A collisions
reach a different initial temperature ($T$), energy density ($\epsilon_i$)
 and $\mu_B$, or  deviate from equilibrium, rendering their comparison inadequate.

\noindent
A
 new parameter adequate to compare strangeness in different colliding systems, the initial energy density
($\epsilon_i$) has been introduced.
This is the
 only parameter characterizing the initial state of the collisions which can be estimated
from the data.

\noindent
We consider in the following only systems which agree with a grandcanonical describtion
e.g. mainly central A+A collisions but also 
elementary collisions like $p \overline{p}$ collisions.
\\

\noindent
Studying the energy dependence of the total $ s \overline{s}$
 production through the strangeness suppression factor $\lambda_s$
 we find two dependences which overlapp each other.
The one is the dependance of $\lambda_s$ on $\mu_b$.
The other is the dependence of $\lambda_s$ on $T$.
\\
After eliminating trivial sources of strangeness
enhancement, like comparing systems with different $\mu_B$ and  $\epsilon_i$
we arrive at a universal behaviour of $\lambda_s$ in central A+A and particle 
(e.g. $p \overline{p}$) collisions. 

\noindent
In particular  $\lambda_s$  at $\mu_B=0$ is simply following the temperature
 rising as a function of $\epsilon_i$
and saturating above $\epsilon_i$ $\sim$ 0.6-1 GeV/fm3.
This is plausible, as  $\lambda_s$ is one of the particle ratios
which enters the very determination of T, and is also strongly T dependant.
It is  more sensitive to $\mu_B$
as the T which is estimated through several ratios, some of them less
sensitive to $\mu_B$ than the $\lambda_s$ ratio.
\\

\noindent
Strangeness (and temperature) enhancement is therefore observed in a dramatic way 
for all systems reaching an  $\epsilon_i$ greater than 0.6-1 GeV/fm3
as compared to all systems with smaller  $\epsilon_i$.
This indicates the onset of the phase transition at 0.6-1 GeV/$fm^3 $
 in
agreement with the $\epsilon_{i}$(crit) predicted by
 lattice QCD estimates.
 The latest fit to data gives  $\epsilon_{i}$(crit) $\sim$ 0.6 $\pm$ 0.2 $\pm$ 0.3 (syst) GeV/$fm^3$.
More data will allow for a more precise estimate of $\epsilon_{i}$(crit).
\\
Note that all above observations are made using hadrons with
 u,d,s quarks and their antiquarks. We therefore can speak about a light flavoured QGP.
The outstanding feature of strangeness production
taken together with light flavoured hadrons is 
that they  allow to estimate the critical parameters,
unlike hard probes like $c \overline{c}$,  $b \overline{b}$ states suppression,
and jet quenching which may be setting in above $T_c$.
Correlation of onsets seen in other features of light flavoured hadrons like flow
\cite{horst_flow},
study of spectral characteristics \cite{kodama},
or fluctuations as a function of energy will enhance the accuracy of this estimate.
\\
\noindent
On the other side, if one argues that 
 the dramatic  rise and saturation of T (and $\lambda_s$) after $\epsilon$ $\sim$ 0.6-1 GeV,
has nothing to do with the QCD phase transition,
also  strangeness enhancement has  nothing to do with  the QCD phase transition.
\\

\noindent
It is an ongoing discussion in the literature  what the saturation of the chemical freeze out T
with increasing energy really means and especially
 why is the same for p+p and A+A
collisions e.g. \cite{satz_hagedorn,dokshitzer,stock_review}.
\\
\noindent
It is  always taken for granted that no QCD phase transition
can appear in a p+p system due to the small volume.
 What if the p+p system has infinite energy ? 
After which energy the initially colliding particle volumes
plays no role anymore ?
\\
\noindent
Maybe it is  not a completely unexpected feature, if the
 hadronization of any quark and gluon system into hadrons
 (e.g. jet hadronization), 
as well as the hadronic mass spectrum, 
do reflect the existance and value of  the $T_c$ of the QCD phase transition.
\noindent
 A "QCD phase transition" in a high multiplicity $p \overline{p}$ collision ?
\cite{gutay}
Along a hadronizing jet ?  \cite{hadronizing}.
\\
While $q \overline{q}$ produced in an elementary  collision
fly apart and cannot communicate with each other \cite{dokshitzer},
the hadronization along each jet and the resulting particle yields may exhibit 
equilibrium features reflecting the $T_c$.  
Maybe it is the QCD vacuum itself which has a thermostat-like nature  \cite{dokshitzer}.
\\
\noindent
However, $p \overline{p}$ collisions, like A+A collisions too,
need a centrality trigger to result in grandcanonical particle ratios
e.g. at midrapidity.

\noindent
How can a high multiplicity $p \overline{p}$ collision appear to be an equilibrated system?
We elaborate on the consequences of quantum mechanical coherence
 for  multiparticle production \cite{dokshitzer}  \cite{stock_review}.
\\
\noindent
High energy particle collisions
should not be viewed as a classical  billiard ball cascade.
Interactions among particles and virtual particle exchange
for a  quantum mechanical coherent system may
lead faster to equilibrium.
\\

\noindent
Figure \ref{ls_1}, right, shows evidence of a universal behaviour
of $\lambda_s$ in both A+A and $p \overline{p}$ collisions.
The same universality is seen for the temperature (fig. \ref{ls_4}).
\\
One could  make these  plots and extract $T_c$
using simply high multiplicity $p \overline{p}$ collisions ?
This is to be proven.
We see however already the same tendency in the N+N data in fig. \ref{ogilvie}, right,
even if these data have not been tested against thermodynamic behaviour.
\\
Evidently, A+A collisions offer  additional crucial means to learn about the QCD
phase transition through  phenomena like $c \overline{c}$, $b \overline{b}$
 suppression, jet quenching, flow effects, fluctuations etc.
Flow of strange particles seem to agree with production through quark coalescence.

\noindent
It appears that many open problems
 need to be clarified in order to arrive in a clear all encompassing
 picture of the QCD phase transition in nature.
Future precision measurements of  hadrons
 at RHIC and the LHC in both A+A and p+p collisions
 will considerably improve
our understanding of strangeness and its role in the QCD phase transition.
Questions about deviations from 
equilibrium will be better addressed through presence of precise and abundant data.
RHIC and LHC data from A+A as well as p+p and p+A collisions,
and
low energy measurements around 30 A GeV Pb+Pb at SPS and the future GSI will be crucial 
for 
measuring the critical $T$ 
with a high accuracy, possibly learn about the order of the transition
and other characteristics
and  detect new critical phenomena like fluctuations.

\subsection{Pentaquarks}

\noindent
The  theoretical description of pentaquarks is advancing very rapidly, however it
 does not lead to a unique picture on the pentaquark existence and characteristics,
reflecting the complexity of the subject.
For example, different lattice calculations give very different results to the questions
if pentaquarks exist  and which mass and parity they have.
The narrow width of $\theta^+$ remains to be understood, as well as it's production
mechanism and $\sqrt{s}$ dependence.
The revival of the question on pentaquark existence is however very 
recent, and a fast progress is expected from both theory and experiment
in the near future.
\\
Many experiments observed the $\theta^+$ ($uudd \overline{s}$)
 peak in different reactions and energies
and
 in two decay channels namely into $n K^+$ and into $p K^0_s$.
The
 systematic errors of each channel differ as well as the backgrounds.
The mass of $\theta^+$ from  $n K^+$ measurements is systematically
higher
 than the mass from  $p K^0_s$, which may be due to Fermi-motion corrections and
the details of missing mass analysis as compared to direct measurements.
If
 we add a systematic error of 0.5\% of the measured mass (therefore
of about 8 MeV) on all measurements for which  no systematic error
was given by the experiments,
we find that
the $\theta^+$ mass from $\theta^+ \rightarrow p K^0_s$
deviate from their
mean mass of 1.529 $\pm$ 0.022 GeV
with a $\chi^2/DOF$ of 0.95.
The $\theta^+$ mass from $\theta^+ \rightarrow n K^+$
for which the systematic errors are given,
gives a mean mass of 1.540 $\pm$ 0.022 GeV
and a $\chi^2/DOF$=0.91.
All observations together give a mean $\theta^+$ mass of 
1.533 $\pm$ 0.031 GeV and they
deviate from their mean with a
 $\chi^2/DOF$
of 2.1, reflecting mainly the difference of masses between
the two considered decay channels.
It is important to understand the origin of this discrepancy.
\\

\noindent
The largest individual $S/ \sqrt{B}$ was 7.8. 
This suggests that an increase of significance is needed to make this result more reliable.
New data taken by the CLAS and LEPS experiments
in 2004, confirm in a preliminary analysis the $\theta^+$
peak, suggesting that the hypothesis of a statistical fluctuation
is not probable.
The possibility of systematic errors need to be excluded. 

\noindent
Neutrino experiments (which have seen the $\theta^+$)
 due to their full acceptance 
and the clean reaction should be more free 
 of bias of the type of kinematic reflexions
than restricted acceptance and/or high multiplicity experiments.

\noindent
The antiparticle
 $\overline{ \theta}^-$ has also been observed with
mass and width consistent with the $\theta^+$ peak by Zeus.

\noindent
Most of the experiments measure a $\theta^+$ width consistent with the experimental resolution,
while few of them give a measurement of width somewhat larger than their resolution
namely Zeus and Hermes.
A measurement with a much improved resolution would be important.
Non-observation of $\theta^+$ in previously taken experimental data lead to an estimate
of its width to be of the order of 1 MeV or less \cite{1mev_arndt}.
This limit would gain in significance, once the $\theta^+$
non-observation by several experiments will be better understood.

\noindent
Preliminary CLAS results exhibit a second peak in the $n K^+$ spectrum
suggesting an excited state of $\theta^+$.
In addition, there is evidence that the $\theta^+$ is preferably produced
by the decay of a new resonance $N^0(2400)$.
There are also preliminary hints for a peak in the $p K^+$
and $\overline{p} K^-$ invariant masses by STAR and CLAS,
however at a different mass of 1.53 and 1.57 GeV.
\\

\noindent
The NA49 experiment observed narrow candidates for the
$\Xi^{--}$(1860) ($ddss \overline{u}$), $\Xi^0(1860)$ ($uuss \overline{u}$)
 and $\Xi^-(1850)$ ($dssu \overline{u}$)
pentaquarks and  for the antiparticles of the first two.
A
 detailed discussion of $\theta^+$ and the $\Xi^{--}$ (called $\phi(1860)$)
have been added recently in the Review of Particle Physics
under exotic baryons \cite{pdg}.
It is entitled $\theta^+$, a possible exotic baryon resonance and has three stars.
The
 mass of the $\Xi(1860)$ is in agreement with the most recent
estimates of the chiral soliton model unlike earlier work
\cite{ellis}.
\\

\noindent
The H1 experiment has observed a candidate for the first charmed pentaquark
the $\theta^0_c(3099)$ ($uudd \overline{c}$)
with width of 12 $\pm$ 3 MeV
and it's antiparticle.
The mass and width of the particle and it's antiparticle are consistent.
\\

\noindent
The  STAR experiment has observed  a 
$N^0$
($udsd \overline{s}$, $udd u \overline{u}$)
 or $\Xi$ ($uds s \overline{d}$) I=1/2 candidate
  with mass 1734  MeV
in Au+Au reactions at $\sqrt{s}$=200 GeV in
 the $\Lambda K^0_s$ invariant mass,
and  width consistent with the experimental resolution of about 6 MeV.
The GRAAL experiment has reported two narrow candidates for $N$
pentaquarks namely at 1670 in the $\eta n$ invariant mass and at 1727 MeV
in the $\Lambda K^0_s$ and the $\Sigma^- K^+$ invariant masses.
The observation of the 1727 MeV peak in two decay channels 
with same mass and width,
reduces considerably the probability that the peak is due to
a kinematic reflexion or other systematic error, as the sources of systematic errors and
the background is  different for the two decay channels studied.
The mass of the peaks at 1670 and (1734, 1727) MeV are in  good agreement with the
$N$ mass of $\sim$
1670, 1730 MeV suggested by a modified partial wave analysis of old data by
 Arndt et al \cite{arndt0312126}. 
\\

\noindent
Furthermore, several high statistics 
experiments report the non-observation of pentaquark states
in particular inclusive studies of $e^+ e^-$ collisions, B decays,
and inclusive high energy $p \overline{p}$ and $pA$ reactions.
Recent publications
discuss problems like e.g. the non-observation of these peaks
by previous experiments \cite{wenig} and the possibility that the
observed peaks are kinematic reflections e.g. \cite{dzierba}.
This  source of systematic errors is investigated by the experiments
which observe pentaquark candidates.
As non-observations are concerned 
the characteristics of the $\theta^+$ production mechanism may be one reason
why this state has not been observed in a number of experiments
\cite{karliner,azimov_n2400}.
Non-observations in high statistics $e^+ e^-$ experiments
in B decays
and high energy $p \overline{p}$, $pA$ experiments,
may be due to pentaquark cross section suppression in $e^+e^-$ reactions
as well as at high energy \cite{0406043,titov}.
The increase in statistics planned by several experiments will
allow to test the current observations with a much better significance
and will allow also a better study of systematic errors,
as well as a measurement of pentaquark characteristics like the parity
and the spin.
Furthermore, not
 only the observations of pentaquark candidates but also
the  non-observations  should be published
in order to facilitate a better understanding of the underline physics.

\begin{theacknowledgments}
I wish to thank the organizers of the workshop for their invitation, for the excellent 
organization, the high scientific quality and the pleasant and fruitfull atmosphere
of this unique workshop. 
I would like to thank Prof. M. Chiapparini for his help for
 the preparation of the proceedings.
Furthermore, I wish to thank Prof. T. Kodama and  Prof. P. Minkowski  
for fruitfull discussions.

\end{theacknowledgments}

\bibliographystyle{aipproc}   

\end{document}